\documentclass[11pt, a4paper]{article}

\usepackage{amsthm}
\usepackage{amsmath}
\usepackage{amssymb}
\usepackage{amsfonts}
\usepackage{latexsym}
\usepackage{booktabs}
\usepackage{multirow}
\usepackage{graphicx}
\usepackage{natbib}
\usepackage{array}
\usepackage[left = 2.5cm, right = 2.5cm, top = 3cm, bottom = 3cm]{geometry}
\usepackage[colorlinks,citecolor=black,urlcolor=black]{hyperref}

\newcommand{\vech}{\mathrm{vech}}

\def\diag{{\rm diag}}
\def\tr{{\rm tr}}

\theoremstyle{definition}
\newtheorem{exmp}{Example}[section]

\newtheorem{theorem}{Theorem}

\title{Directional tests in Gaussian graphical models}

\author{Claudia Di Caterina \\ \texttt{claudia.dicaterina@univr.it} \smallskip \\
	Department of Economics, University of Verona \\ 
	37129 Verona, Italy 
	\bigskip \\
	Nancy Reid \\ \texttt{reid@ustat.toronto.edu} \smallskip \\
	Department of Statistical Sciences, University of Toronto \\
	Toronto, Canada M5S 3G3
	\bigskip \\
	Nicola Sartori \\ \texttt{sartori@stat.unipd.it} \smallskip \\
	Department of Statistical Sciences, University of Padova \\
	35121 Padova, Italy \\ 
	\bigskip
}

\begin{document}
	
	\maketitle
	
	\begin{abstract}\noindent
		Directional tests to compare incomplete undirected graphs are developed in the general context of covariance selection for Gaussian graphical models. The exactness of the 
		underlying saddlepoint approximation is proved for chordal graphs and
		leads to exact control of the size of the tests, given that the only approximation error involved is due to the numerical calculation of two scalar integrals. Although exactness is not guaranteed for non-chordal graphs, the ability of the saddlepoint approximation to control the relative error leads the directional test to overperform its competitors even in these cases. The accuracy of our proposal is verified by simulation experiments
		under challenging scenarios, where inference via standard asymptotic approximations to the likelihood ratio test and some of its higher-order modifications fails. 
		The directional approach is
		used to illustrate the assessment of
		Markovian dependencies in a dataset from a veterinary trial on cattle.
		A second example with microarray data shows how to
		select the graph structure related to genetic anomalies due to acute lymphocytic leukemia.
		
		\noindent
		\textit{Keywords:} Covariance Selection; Exponential Family; Higher-order Asymptotics; Likelihood Ratio Test; Saddlepoint Approximation; Undirected Graph.
	\end{abstract}
	
	\section{Introduction}
\label{sec:intro}
Undirected graphical models have gained considerable success in a variety of 
fields, including medicine, social sciences and physics, due to their flexibility and 
easy interpretation. Typically, these probabilistic graphs describe complex multivariate distributions of variables (nodes) through the product of simpler sub-models, each referred to a low-dimensional subset of the graph (clique).
Book-length expositions on the topic can be found in \cite{Lauritzen:1996}, 
\cite{borgelt02},  and \cite{whitta09}. 

Today, applications of graphical models are challenged by  the growth in size and sophistication of modern data. An important question is inferring the structure of large graphs, i.e. the underlying connections (edges) between the variables under examination. This task is well known in the literature by the name of covariance selection. 
A very popular class of graphical models is that of decomposable models, which describe graphs that contain no chordless cycles of length greater than 3. These graphs are called chordal, decomposable or triangulated \citep[Sect. 2.1]{Lauritzen:1996}.

For reasons of convenience, a graphical model is often expressed by means of the exponential family form. The Gaussian distribution is 
particularly suitable for continuous responses, as conditional independence in the graph can be easily characterized in terms of assumptions on model parameters
(see Section \ref{sec:GM1}). 

Likelihood-based inference for covariance selection is discussed in  \citet{salg05} in the context of testing exclusion of single edges in complete graphs, i.e. fully saturated models. 
\citet[Sect. 7]{cordoba20} review general edge exclusion tests, acknowledging the poor quality of the usual chi-squared approximation to the distribution of the likelihood ratio statistic. They mention that, when testing the removal of $r$ edges, the exact distribution is the 
product of $r$ Beta random variables \citep[Prop. 5.14]{Lauritzen:1996}. However, this result has not received much attention in the literature and 
seems of limited practical utility. 
Another strategy consists in carrying out iteratively exclusion tests for single edges based on partial correlation coefficients, with some adjustment needed to account for multiple comparisons.

In this paper we  develop likelihood-based directional tests for covariance selection in Gaussian graphical models, possibly incorporating {\it a priori} restrictions on the graph structure. 	
Specifically, our method allows to test hypotheses that involve removal of  sub-graphs with multiple edges from complete or incomplete graphs.
We prove the exactness of the underlying saddlepoint approximation for chordal graphs and run extensive
Monte Carlo simulations 
which show the null uniform distribution of the directional $p$-value in 
challenging scenarios, even when the number of nodes is 
larger than the sample size. In those settings, the classical approach based on 
the likelihood ratio statistic or some of its higher-order modifications \citep{Skovgaard:2001} breaks down.
We also show  results for a non-chordal graph, where  directional inference is confirmed to be  more accurate than its competitors.
A much simpler problem in covariance selection,  limited to testing an incomplete graph versus the saturated model, was studied by \citet[Sect. 5.3]{dav14} and shown to be exact in \citet{huang21}. Our extension involves both theoretical and computational innovations.

Directional inference on a vector-valued parameter of interest was introduced by \cite{Fraser.Massam:1985} in nonnormal linear regression models 
and then generalized in \cite{skovgaard1988}.
Substantial progress from both a methodological and computational perspective was made by \cite{dav14}, where the computation of the directional $p$-value by one-dimensional numerical integration
proved especially accurate 
in several settings.
The procedure
was extended from linear exponential families to nonlinear parameters of interest in general continuous
models by \citet{fraser16}. Besides its accuracy, the directional approach was found to coincide with exact results in several classical situations \citep{Mccor19}.

Section \ref{sec:background} 
reviews the technique of directional inference for exponential family models.
{Section \ref{sec:GM} presents the new directional testing method for covariance selection; this involves  proving the exactness of the saddlepoint approximation for decomposable Gaussian graphical models in chordal graphs 
	and developing specific notation also valid in the non-chordal case. A
	number of computational innovations can then be found in Section \ref{sec:com}.
	Simulation results comparing the accuracy of the various methods are shown in Section \ref{sec:sim}, while Section \ref{sec:app} reports applications to data from a veterinary trial and from a microarray study of altered gene expression in acute lymphocytic leukemia. Comments and final remarks are made in Section \ref{sec:disc}.

	\section{Background}\label{sec:background}
	\subsection{Likelihood ratio tests}\label{sec:lrt}
	Assume that 
	$y$ follows a parametric distribution $f(y;\theta)$, with $\theta\in \mathbb{R}^p$. 
	The log-likelihood function $\ell(\theta)=\ell(\theta;y)=\log f(y;\theta)$ is maximized by the maximum likelihood (ML) estimator $\hat\theta=\hat{\theta}(y)$. Whenever appropriate, the notation $\ell^0(\theta)=\ell(\theta;y^0)$ and $\hat\theta^0=\hat{\theta}(y^0)$ will be adopted to stress the dependence of those quantities on the observed data point $y^0$.
	Possibly after a reparameterization, the model parameter can be typically expressed as $\theta=(\psi, \lambda)$, where $\psi(\theta)$ is the $d$-dimensional component of interest involved in the hypothesis $H_\psi\!: \psi(\theta)=\psi$. We write $\hat\theta_\psi=(\psi, \hat\lambda_\psi)$ to denote the constrained ML estimator of $\theta$ under $H_\psi$.
	
	Under usual regularity conditions  \cite[see, e.g.,][Sect. 9.3]{Cox.Hinkley:1974}, the first-order approximation to the distribution of $\hat\theta$ is normal with mean $\theta$ and estimated covariance matrix $j(\hat\theta)^{-1}$, with $j(\theta)=-\partial^2\ell(\theta)/\partial\theta\partial\theta^\top$ the observed Fisher information matrix. 
	The hypothesis $H_\psi$ can be tested via the likelihood ratio statistic
	\begin{equation}\label{lrt}
		w(\psi) = 2\{\ell(\hat\theta) - \ell(\hat\theta_\psi)\}\,,
	\end{equation}
	which 
	{is invariant to reparameterizations} and has an approximate $\chi^2_d$ distribution under the null hypothesis $H_\psi$, $d$ being the 
	dimension of the parameter of interest $\psi$.
	
	\cite{Skovgaard:2001} introduced two modifications to (\ref{lrt}), namely
	\begin{equation}\label{skov}
		w^*(\psi) = w(\psi) \bigg\lbrace 1-\frac{\log\gamma(\psi)}{w(\psi)}\bigg\rbrace ^2 \quad \text{and} \quad
		w^{**}(\psi) = w(\psi) -2\log\gamma(\psi)\,,
	\end{equation}
	and showed that the limiting distribution of 
	both test statistics based on the correction factor $\gamma(\psi)$ is also $\chi^2_d$. These modifications were obtained by analogy with the derivation for scalar parameters of interest of modifications to the square root of $w(\psi)$, the so-called $r^*$ approximation of \cite{barn86}, further discussed in \citet{fraser99}. 
	\cite{Skovgaard:2001} emphasized not only the simplicity of computation of the adjustment, especially when compared to \cite{bartlett37} correction using moments, but also its large-deviation properties.
	
	Tests based on $w(\psi)$, including $w^*(\psi)$, $w^{**}(\psi)$ and the Bartlett-corrected $w(\psi)$, provide omnibus measures of departure of the data from $H_\psi$: the resulting $p$-value averages the deviations from the null hypothesis in all potential directions of the parameter space. In the next section, the approach of \citet[Sect. 3]{dav14} for measuring the departure from $H_\psi$ only in the direction indicated by the observed data will be reviewed. For a more complete exposition of the difference between omnibus and directional tests, see \cite{Fraser.Reid:2006a}.

	\subsection{Directional tests in linear exponential families}
	\label{sec:dir}
	Focusing on hypotheses that are linear in the canonical parameter $\theta$ of an exponential family model,
	we shall summarize here the procedure detailed in \citet[Sect. 3]{dav14} which involves two steps of dimensionality reduction. 
	
	Denoting by $u=u(y)$ the sufficient statistic for the $p$-dimensional vector parameter $\theta$, 
	we can consider
	the marginal density of $u$ and the corresponding 
	log-likelihood function $\ell(\theta;u) = \theta^\top u - K(\theta)$, which takes the standard exponential family form.
	{{Consistent with the notation established by \cite{dav14} and \cite{fraser16}, we  define the observed data $y^0=(y^0_1,\ldots,y^0_n)$  and the corresponding observed value of the sufficient statistic $u^0 = u(y^0)$.}
		Given
		the centered statistic $s = u - u^0$ with observed value $s^0=u^0 - u^0=0$,} the tilted log-likelihood function is
	\begin{equation}\label{likscore}
		\ell(\theta;s)   =  \theta^\top s + \ell^0(\theta) \,,
	\end{equation}
	where   $\ell^0(\theta) =  \ell(\theta; u=u^0)$.
	
	When linearity in $\theta$ applies to both the interest and nuisance parameters, meaning $\theta = (\psi, \lambda)$, expression (\ref{likscore}) can be written as
	\begin{equation}\label{explik}
		\ell(\theta;s) = \psi^\top s_1 + \lambda^\top s_2 + \ell^0(\psi,\lambda)\,,
	\end{equation}
	where $\psi$ and $s_1$ have dimension $d$.
	The first dimensionality reduction from $p$ to $d$
	follows directly from conditioning on the component of the statistic sufficient for $\lambda$. Indeed, 
	the conditional distribution of $s_1$ given $s_2$ depends on $\psi$ only 	{ and is still of exponential family form 
		{\citep[cf.][Lemma 2.7.2]{lehmann05}}}.
	Such a conditioning translates into fixing 
	$\hat\theta_\psi = (\psi, \hat\lambda_\psi)$ at the observed value $\hat\theta^0_\psi= (\psi, \hat\lambda^0_\psi)$. 
	
	The saddlepoint approximation for this conditional distribution  is typically very accurate \citep{Barn79}. {Following for instance \citet[Sect. 10.10.2]{ps97}, we can illustrate how the saddlepoint approximation is obtained as the ratio of the saddlepoint approximation for the joint density of $s=(s_1,s_2)$ and the saddlepoint approximation for the marginal density of $s_2$. Indeed, the former can be expressed as
		\begin{equation}\label{saddlejoint}
			\dfrac{\exp[\{\theta-\hat\theta(s)\}^\top s + \ell^0(\theta) - \ell^0\{\hat\theta(s)\}]}{(2\pi)^{p/2}|-\ell^0_{\theta\theta}\{\hat\theta(s)\}|^{1/2}} = \dfrac{\exp[\ell(\theta;s) - \ell\{\hat\theta(s);s\}]}{(2\pi)^{p/2}|j_{\theta\theta}\{\hat\theta(s)\}|^{1/2}}\,,
		\end{equation}
		where $\hat\theta(s)$ solves in $\theta$ the score equation from the log-likelihood (\ref{explik}), $s = -\ell^0_{\theta}(\theta) = - \partial\ell^0(\theta)/\partial\theta$, and
		${j_{\theta\theta}(\theta)} = - \partial^2\ell(\theta;s)/\partial \theta\partial \theta^\top = - \partial^2\ell^0(\theta)/\partial \theta\partial \theta^\top=-\ell^0_{\theta\theta}(\theta)$. Similarly, the saddlepoint approximation for the marginal distribution of $s_2$ is
		\begin{equation}\label{saddlemarg}
			\dfrac{\exp[\{\lambda-\hat\lambda_\psi(s_2)\}^\top s_2 + \ell^0(\theta) - \ell^0\{\hat\theta_\psi(s_2)\}]}{(2\pi)^{(p-d)/2}|-\ell^0_{\lambda\lambda}\{\hat\theta_\psi(s_2)\}|^{1/2}} = \dfrac{\exp[\ell(\theta;s) - \ell\{\hat\theta_\psi(s_2);s\}]}{(2\pi)^{(p-d)/2}|j_{\lambda\lambda}\{\hat\theta_\psi(s_2)\}|^{1/2}}\,,
		\end{equation}		
		where $\hat\theta_\psi(s_2)=(\psi,\hat\lambda_\psi(s_2))$ is the solution to the score equation from the log-likelihood (\ref{explik}), seen as a function of $\lambda$ for fixed $\psi$, $s_2 = -\ell^0_{\lambda}(\theta) = - \partial\ell^0(\theta)/\partial\lambda$, and
		${j_{\lambda\lambda}(\theta)} = - \partial^2\ell(\theta;s)/\partial \lambda\partial\lambda^\top = - \partial^2\ell^0(\theta)/\partial \lambda\partial \lambda^\top=-\ell^0_{\lambda\lambda}(\theta)$.
		The ratio of (\ref{saddlejoint}) and (\ref{saddlemarg}) when $s_2=0$ gives the following saddlepoint approximation for the density of $s_1$ given $s_2=0$, also called double saddlepoint approximation, for the reduced model in $\mathbb{R}^d$: \begin{equation}\label{conddens}
			h(s;\psi) = c \exp[\ell(\hat\theta_\psi^0;s) - \ell\{\hat\theta(s);s\}]\, {|j_{\theta\theta}\{\hat\theta(s)\}|}^{-1/2} \,, \quad s \in {\cal L}^0\,, 
		\end{equation}
		where the normalizing constant $c$ includes all factors not depending on $s_1$, and ${\cal L}^0$ is the $d$-dimensional plane described
		by setting 
		$s_2=0$, or equivalently 
		$\hat\theta_\psi = \hat\theta^0_\psi$.    
		The relative error of the approximation (\ref{conddens}) is typically of order $O(n^{-1})$, with $n$ number of independent observations, but it reduces to $O(n^{-3/2})$ after re-normalization. For a comprehensive review of saddlepoint approximations and their statistical applications, see \citet{butler07}.}
	The following example with scalar parameter of interest ($d=1$) illustrates the use of the tilted log-likelihood function (\ref{explik}) in the derivation of the saddlepoint approximation (\ref{conddens}).
	\begin{exmp}[Univariate normal distribution]
		{Let $y_1,\ldots,y_n$ be a random sample from a $N(\mu,\sigma^{2})$ distribution.
			The log-likelihood function in exponential family form is
			\begin{equation*}\label{lltheta}
				\ell(\theta)=	\ell(\psi,\lambda)=\psi u_1 + \lambda u_2 + \dfrac{n}{2}\log\left(-2\psi\right) + \dfrac{n\lambda^2}{4\psi},
			\end{equation*}
			where  $\theta=(\psi,\lambda)=(-1/2\sigma^2, \mu/\sigma^2)$ is the canonical parameter and $u=(u_1,u_2)=(\sum_i y_i^2,$ $\sum_i y_i)$ is the minimal sufficient statistic with observed value $u^0=(u^{0}_1,u^0_2)$.
			The tilted log-likelihood (\ref{explik}) expressed as a function of the centered sufficient statistic $s=u - u^0$
			is
			\begin{equation*} \label{llthetaTilt}
				\ell(\theta;s)=	\ell(\psi,\lambda; s) =\psi (s_1+u_1^0) + \lambda (s_2+u_2^0) + \dfrac{n}{2}\log\left(-2\psi\right) + \dfrac{n\lambda^2}{4\psi}.
			\end{equation*}
			After some algebra, the unnormalized saddlepoint approximation (\ref{conddens}) in $\mathcal{L}^0 = \{(s_1,s_2): s_1 > 
			- u_1^0 + (u_2^0)^2/n, 
			s_2=0\}$ can be written as
			\begin{equation}\label{conddens00}
				h(s;\psi) \propto \exp\left\{\psi \left[
				s_1+u_1^0 - \dfrac{(u_2^0)^2}{n}
				\right]\right\}  
				\left\{
				s_1+u_1^0 - \dfrac{(u_2^0)^2}{n}
				\right\}^{\frac{(n-1)}{2}-1}, 
			\end{equation}
			where 
			$u_1^0 - (u_2^0)^2/n$ is $n$ times 
			the unadjusted sample variance. 
			In this simple case, the saddlepoint approximation is exact:
			(\ref{conddens00}) coincides with the kernel of a $\chi^2_{n-1}/(-2\psi)$ distribution, which is the exact conditional distribution of $s_1=u_1-u_1^0$ given $s_2=u_2-u_2^0=0$. This is consistent with the more general result in \citet{Mccor19}.
		}
	\end{exmp}
	
	{ The second dimensionality reduction from $d$ to 1, not needed in the previous example, }consists of constructing
	a one-dimensional conditional distribution for $s$ along the direction indicated by the data. 
	With this aim, denote by $s_\psi$ the expectation of $s$ under model (\ref{conddens}) if $H_\psi$ holds, that is the value of $s$ for which $\theta = \hat\theta_\psi^0$ is the constrained ML estimate:
	\begin{equation}\label{spsi}
		s_\psi = - \ell_{\theta}^0(\hat\theta_\psi^0)= \begin{pmatrix} -\ell^0_\psi(\hat\theta_\psi^0)\\ 0 \end{pmatrix},
	\end{equation}
	depending on the observed data point $y^0$.
	The line ${\cal L}^*$, in ${\cal L}^0$, which joins the observed value $s^0=0$ and the expected value $s_\psi$
	can be parameterized by a scalar $t \in \mathbb{R}$:
	$$
	s(t) = s_\psi + t(s^0-s_\psi) = (1-t)s_\psi\,,
	$$
	and consequently the maximum likelihood estimate $\hat\theta(s)$ in (\ref{conddens}) can vary with $s(t)$. 
	The approximation (\ref{conddens}) constrained to ${\cal L}^*$ is used to compute the $p$-value, the probability that $s(t)$ is as far or farther from $s_\psi$ than is the observed value $s^0=0$.
	The directional $p$-value which measures the deviation from $H_\psi$ along the line ${\cal L}^*$ is thus 
	\begin{equation}\label{pval}
		p(\psi) = \frac{\int_1^{t_{\sup}} t^{d-1}h\{s(t);\psi\}\, \mathrm{d}t}{\int_0^{t_{\sup}} t^{d-1}h\{s(t);\psi\}\, \mathrm{d}t}\,,
	\end{equation}
	where $t=0$  and $t=1$ correspond respectively to  $s=s_\psi$ and to the observed value $s^0=0$.   
	The factor $t^{d-1}$ is due to the Jacobian of the
	transformation from the variable { $s\in {\cal L}^0 $} to { polar coordinates} $(\|s\|, s/\|s\|)$ \citep[Sect. 3.2]{dav14}. 
	The upper limit of the integrals in (\ref{pval}) is the largest value of $t$ for which the ML estimator corresponding to $s(t)$ exists, and in some situations can be determined analytically. The directional $p$-value in one dimension gives the probability to the right of the observed value, conditional on the observed value being to the right of the expected value under $H_\psi$, i.e. the probability in the right tail of the distribution. In higher dimensions the $p$-value is the probability of being `further out' on the line connecting the expected value under $H_\psi$ to the observed value, conditional on being on that line \citep[Sect. 2]{dav14}.
	
	
	As  in \citet[Sect. 3.2]{dav14},  the relative error of formula (\ref{pval}) inherits that of the  saddlepoint approximation  (\ref{conddens}) {after  re-normalization}, so is typically $O(n^{-3/2})$ in continuous models.
	When the re-normalized saddlepoint approximation is exact,  then the directional test will also be exact, as the re-normalization is automatically incorporated in (\ref{pval}). \cite{Mccor19} established this exactness for a number of tests for multivariate normal models, and \citet{huang21} were able to prove exactness for the case of testing a saturated model in \citet[Sect. 5.3]{dav14}. The exactness in our setting is shown in Section \ref{sec:dirp} for chordal graphs. In addition,  numerical results in the last simulation scenario of Section \ref{sec:sim} illustrate the extreme accuracy of the directional approach even in situations where
	the alternative graph is non-chordal. 

	Using the notation established in this section,
	we also give the form of the term $\gamma(\psi)$ appearing in (\ref{skov}) under exponential family models. Specifically, equation (13) in \cite{Skovgaard:2001} is
	\begin{equation}\label{gamma-supp}
		\gamma(\psi) = \frac{\{(s - s_\psi)^\top j_{\theta\theta}^{-1}(\hat\theta_\psi)(s - s_\psi)\}^{d/2}}{w^{d/2-1} (\hat\theta - \hat\theta_\psi)^\top(s - s_\psi)}
		\left\{\frac{|j_{\theta\theta}(\hat\theta_\psi)|}
		{|j_{\theta\theta}(\hat\theta)|}\right \}^{1/2}\,,
	\end{equation}
	to be evaluated at $s=0$ when computing the corresponding observed $p$-value.

	\section{Directional tests for Gaussian graphical models}
	\label{sec:GM}
	\subsection{Notation and setup}\label{sec:GM1}
	Gaussian graphical models are very useful for describing normal multivariate distributions using the nodes and edges of a related graph. The nodes correspond to variables and the lack of an edge between two nodes models the conditional independence of the two variables, given the remaining ones. This corresponds to a zero entry in the concentration (inverse covariance) matrix \citep{Lauritzen:1996}, and covariance selection involves identifying these conditional independencies.
	
	Let $y_1,\ldots,y_n$ be a random sample from the $q$-variate normal distribution $N_q(\mu,\Omega^{-1})$, where the mean is $\mu \in \mathbb{R}^q$ and the $q \times q$ concentration matrix $\Omega$ is positive definite. The log-likelihood {function} for $(\mu,\Omega)$ is
	\begin{equation}\label{llmu}
		\ell(\mu,\Omega; y)=\frac{n}{2}\log |\Omega| - \frac{1}{2} \tr(\Omega y^\top y)+1_n^\top y\,\Omega  \mu - \frac{n}{2} \mu^\top \Omega \mu\,,
	\end{equation}
	where $y$ denotes the $n \times q$ matrix with $l$th row vector $y_l^\top$ and $1_n$ is a $n\times 1$ vector of ones. The  ML estimates of $\mu$ and $\Omega$  are 
	$$
	\hat\mu=y^\top 1_n/n, \qquad \hat\Omega= (y^\top y/n - y^\top 1_n 1_n^\top y/n^2)^{-1}\,.
	$$
	
	For covariance selection the mean parameter is not of direct interest, so we focus instead on the marginal distribution of the ML estimator for the covariance matrix $\hat\Omega^{-1}\sim W_q(n-1, \Omega^{-1}/n) $, where $W_q$ denotes the Wishart random variable of order $q$. The marginal log-likelihood function for $\Omega$
	\[ 
	\ell(\Omega; y) = \dfrac{n - 1}{2}\log |\Omega| - \dfrac{n}{2} \tr(\Omega\hat\Omega^{-1})\,,
	\]
	sometimes referred to as restricted log-likelihood or REML, can then be used to carry out inference just on the concentration matrix.
	The directional $p$-value for testing constraints on $\Omega$ in  Section \ref{sec:dirp} is equal to that obtained from the full log-likelihood function (\ref{llmu}), because of the independence between $\hat \mu$ and $\hat\Omega$.
	It is also convenient to exploit the symmetry of the concentration matrix and express the restricted log-likelihood as
	\begin{equation}\label{lll}
		\ell(\omega; u) = \dfrac{n - 1}{2}\log |\Omega| - \dfrac{n - 1}{2} \omega^\top J u\,,
	\end{equation}
	where $\omega = \mathrm{vech}\,\Omega$, $u = 
	n/(n - 1)\mathrm{vech} \,\hat\Omega^{-1}$ and the matrix $J = G^\top G$ is diagonal with elements equal to either 1 or 2. If $A$ is a $q\times q$ symmetric matrix, $\mathrm{vec}\,A$ is the $q^2\times 1$ vector which stacks the columns of $A$ on top of one another, while $\mathrm{vech}\,A$ retains only the $q^* = q(q+1)/2$ entries in the lower triangle of $A$. The two vectors are linked by the relationship $ \mathrm{vec}\,A = G\,\mathrm{vech}\,A$, which also gives the $q^2\times q^*$ duplication matrix $G$ \citep[see, e.g.,][Sect. 11.3]{abadir05}.
	
	In the saturated case addressed by \citet[Sect. 5.3]{dav14}, i.e. the case of a complete graph where $\Omega$ has no particular a priori structure, the condition $n > q$ is required for the existence of $\hat\Omega$ \citep[Theorem 5.1]{Lauritzen:1996}. On the other hand, if the graph is incomplete with some zero off-diagonal entries in $\Omega$,  the ML estimate exists 
	if $n$ is larger than the maximal clique size of the hypothesized graph or its decomposable version \citep[Sect. 5.3.2]{buhl93, Lauritzen:1996}.
	In what follows, we focus on comparing nested unsaturated models corresponding to nested incomplete graphs. Therefore we allow the sample size $n$ to be smaller than the number of nodes $q$, but large enough for the ML estimate of the concentration matrix to exist under the alternative model under study (cf. Section \ref{nonsat}).

	\subsection{Likelihood quantities for unsaturated models} \label{nonsat}
	Suppose some off-diagonal elements $\Omega_{ij}, 1\leq i<j \leq q$, in the concentration matrix are known to be zero, meaning that the underlying graph is known to be incomplete.
	As in \cite{rov96}, we can rearrange the elements of $\omega, u$ and the leading diagonal of $J$ to simplify the calculations. Specifically, defining the edge sets
	\begin{align} \label{edgeset}
		k =  \{ (i,j): \Omega_{ij} \neq 0, i \leq j\}  \quad \text{and} \quad h = \{ (i,j):\Omega_{ij} = 0, i < j \}\,, 
	\end{align}
	and giving any ordering to $k$ and $h$ such that
	\begin{align*}
		k &= \{ k_1, k_2, \dots, k_p \} \quad \text{and} \quad h = \{ h_1, h_2, \dots, h_w \}\,,
	\end{align*}
	it is possible to define
	\[ 
	\omega = \begin{pmatrix}
		\omega_k\\
		\omega_h
	\end{pmatrix}\,,\quad
	u = \begin{pmatrix}
		u_k\\
		u_h
	\end{pmatrix}\,,\quad
	J = \begin{pmatrix}
		J_{kk} & 0\\
		0 & J_{hh}
	\end{pmatrix}\,.
	\]
	Since in unsaturated models $\omega_h = 0$, we can write $\Omega = \Omega_k=\Omega(\omega_k)$ so that the log-likelihood (\ref{lll}) becomes 
	\begin{equation}\label{lllk}
		\ell(\omega_k;u_k) = \dfrac{n - 1}{2}\log |\Omega_k| - \dfrac{n - 1}{2} \omega_k^\top J_{kk} u_k\,,
	\end{equation}
	which is a function of the $p$-dimensional canonical parameter $\theta =\omega_k$ only, with $p> q$. Differentiation of (\ref{lllk}) with respect to $\omega_k$ leads to the score function
	\[ 
	\ell_{\omega_k}(\omega_k) = \dfrac{n - 1}{2} J_{kk} (\sigma_k - u_k)\,,
	\]
	where $\sigma_k$ is the partition of $\sigma=\mathrm{vech}\,\Omega_k^{-1}$ 
	obtained according to (\ref{edgeset}).
	Solving the score equation leads to $\hat\sigma_k=u_k$ and to the corresponding ML estimate $\hat\omega_k$, usually derived numerically \citep[see][Sect. 5.3]{dav14}.
	
	As the observed and expected information matrices are equal in canonical exponential families, from the results in \citet[Sect. 3]{rov96} follows that 
	\begin{equation}\label{jkk}
		j_{\omega_k\omega_k} (\omega_k)= 
		\dfrac{n - 1}{4}J_{kk} \mathrm{Iss}(\Omega_k^{-1})_{kk}J_{kk}\,,
	\end{equation}
	where $\mathrm{Iss}(\Omega_k^{-1})_{kk}$ is a $p \times p$ partition of the 
	Isserlis matrix of the covariance matrix $\Sigma=\Omega_k^{-1}$ \citep{iss18}. 
	The entries of $\mathrm{Iss}(\Sigma)_{kk}$ are 
	\[ 
	\text{Cov}(u_{ij}, u_{rs})=\Sigma_{ir}\Sigma_{js}+\Sigma_{is}\Sigma_{jr}\,,
	\]
	with $(i,j), (r,s) \in k$. 
	
	\subsection{Comparison of nested unsaturated models}\label{sec:dirp}
	Consider now the partition $\omega_k=(\psi, \lambda)$ of the canonical parameter, where $\psi$ is the component of interest having dimension $d\leq p-q$. The null hypothesis $H_0: \psi = \psi_0=0$ tests whether $d$ additional off-diagonal entries $\Omega_{ij}, i<j,$ are zero.
	Hence, the reduced null model is nested in the alternative unsaturated model of Section  \ref{nonsat}. Starting from (\ref{lllk}), the log-likelihood ratio statistic for testing $H_0$ is
	\begin{equation}\label{lrtcovselect}
		w(\psi_0) = -(n-1)\log|\hat\Omega_k^{-1}\hat\Omega_0|\,,
	\end{equation}
	where $\hat\Omega_k=\Omega(\hat\omega_k)$ is the ML estimate  of $\Omega$ obtained from (\ref{lllk}), and $\hat\Omega_0=\Omega(\hat\omega_{k0})$ is its constrained ML estimate under $H_0$, with $\hat\omega_{k0}=(0, \hat\lambda_0)$. The null asymptotic distribution of $w(\psi_0)$  is $\chi^2_d$,  assuming $p$ and $d$ fixed with $n$ that goes to infinity.
	
	For the directional $p$-value that discriminates between two nested Gaussian graphical models, as specified in (\ref{spsi}) we first find the expected value of $s$ under $H_0$
	\begin{equation*}
		s_{\psi_0}= - \ell_{\omega_k}(\hat\omega_{k0})= \dfrac{n - 1}{2} J_{kk} (u_k - \hat \sigma_{k0})\,,
	\end{equation*}
	where $\hat \sigma_{k0}=\vech \,\hat\Omega_0^{-1}$. Then, the log-likelihood function (\ref{likscore}) along the line $s(t)=(1-t)s_{\psi_0}$  follows from (\ref{lllk}):
	\begin{equation}\label{tiltll}
		\ell\{\omega_k;s(t)\}=\dfrac{n - 1}{2}\log |\Omega_k| - \dfrac{n - 1}{2} \omega_k^\top J_{kk} \{ \hat\sigma_{k0} + t(u_k - \hat\sigma_{k0})\}\,.
	\end{equation}
	The maximization of (\ref{tiltll}) entails that $\hat\sigma_{k}\{s(t)\} =\hat\sigma_{k}(t) = \hat\sigma_{k0} + t (u_k - \hat\sigma_{k0})$ or, equivalently, 
	\begin{equation}\label{invlam}
		\hat\Omega^{-1}_k\{s(t)\}=\hat\Omega^{-1}_k(t)=  t \hat\Omega_k^{-1} + (1-t) \hat\Omega_0^{-1}\,.
	\end{equation}
	Given that $\hat\Omega_k(t)=\Omega\{\hat\omega_k(t)\}$, by taking the inverse of the matrix resulting in the left-hand side of  (\ref{invlam})
	the value of $\hat\omega_k(t)$ is obtained accordingly. 
	The replacement of $\omega_k$ in (\ref{tiltll}) with $\hat\omega_k(t)$ and $\hat\omega_{k0}$, respectively, delivers the result
	\begin{align*}
		\exp[\ell\{\hat\omega_{k0};\!s(t)\} \!- \!\ell\{\hat\omega_k(t);\!s(t)\}] &\propto |\hat\Omega_k(t)|^{-\frac{n-1}{2}}\!\exp\!\!\bigg[\!\frac{n-1}{2}\{\hat\omega_k(t)\! -\! \hat\omega_{k0}\}\!^{\top}\!\! J_{kk}\hat\sigma_{k}(t)\!\bigg]\\ 
		&\propto |\hat\Omega_k(t)|^{-\frac{n-1}{2}}\,,
	\end{align*}
	since the function $\{\hat\omega_k(t) - \hat\omega_{k0}\}\!^{\top}\! J_{kk}\hat\sigma_{k}(t)$ is zero (see proof in Appendix \ref{appB}). 
	By (\ref{jkk}), we obtain $|j_{\omega_k\omega_k}(\omega_k)| \propto |\mathrm{Iss}(\Omega_k^{-1})_{kk}|$ and consequently 
	$${ |j_{\omega_k\omega_k}\{\hat\omega_k(t)\}|}^{-1/2} \propto |\mathrm{Iss}\{\hat\Omega_k^{-1}(t)\}_{kk}|^{-1/2}\,.$$
	Thus, following expression (\ref{conddens}), the directional test is based on $p(\psi_0)$ in (\ref{pval}) with
	\begin{equation}\label{hfun}
		h\{s(t); \psi_0\} \propto |\hat\Omega^{-1}_k(t)|^{\frac{n-1}{2}} |\mathrm{Iss}\{\hat\Omega_k^{-1}(t)\}_{kk}|^{-1/2}\,,
	\end{equation}
	and the analytical value of $t_{\sup}$ calculated as detailed in Section \ref*{sec:tsup}.
	If the alternative model were saturated, with $q^*$-vector $\omega_k=\omega$, then
	\begin{equation*}
		|\mathrm{Iss}\{\hat\Omega_k^{-1}(t)\}_{kk}|=|\mathrm{Iss}\{\hat\Omega_k^{-1}(t)\}|=2^q|\hat\Omega_k^{-1}(t)|^{q+1}\,,
	\end{equation*}
	according to the general expression for computing the determinant of the Isserlis matrix \citep[Sect. 2]{rov98}. {In this case (\ref{hfun}) reduces} to
	\begin{equation*}
		h\{s(t); \psi_0\}  \propto |\hat\Omega^{-1}_k(t)|^{\frac{n-1}{2}} |\hat\Omega_k^{-1}(t)|^{-\frac{q+1}{2}} =
		|\hat\Omega^{-1}_k(t)|^{(n-q-2)/2}\,,
	\end{equation*}
	which agrees with the simpler result obtained by \citet[Sect. 5.3]{dav14} for testing the absence of some connections in the complete graph. 
	
	Expression (\ref{hfun}) gives the unnormalized saddlepoint approximation to the distribution of $s(t)$ in ${\cal L}^*$. 
	The following theorem, whose proof is deferred to
	Appendix \ref{appA}, states when (\ref{hfun}) is also the unnormalized exact null conditional density of $s(t)$ in ${\cal L}^*$.
	
	\begin{theorem}\label{th}
		Let $Y \sim N_q(\mu, \Omega^{-1})$ denote a Gaussian graphical model with log-likelihood (\ref{lllk}).
		If the induced incomplete graph is chordal, then (\ref{hfun}) gives the unnormalized exact conditional density of $s(t)$ in ${\cal L}^*$ under $H_0:\psi=\psi_0=0$.
	\end{theorem}
	
	\noindent
	The normalizing constant simplifies in the ratio of integrals in (\ref{pval}), so the approximation error involved in the calculation of the directional $p$-value stems only from the one-dimensional numerical integrations. 
	It is possible to conclude that in Gaussian graphical models describing chordal graphs the saddlepoint approximation to the null conditional density of the sufficient statistic is exact. Consequently, when we test for a reduced graph the resulting directional $p$-value is exactly uniformly distributed under the null hypothesis $H_0:\psi={\psi_0}=0$.

	Monte Carlo experiments in Section \ref{sec:sim} attest this theoretical result, and empirically show that the directional $p$-value stays remarkably accurate in the last simulation scenario based on non-chordal graphs. When the exactness does not hold, indeed, the relative error of the saddlepoint approximation is still of order $O(n^{-3/2})$ as opposed to the absolute error of order $O(n^{-1})$ of the chi-squared approximation to the distribution of $w(\psi_0)$.
	
	Finally, we give the term $\gamma(\psi)$ in (\ref{gamma-supp}) appearing in \citeauthor{Skovgaard:2001}'s (\citeyear{Skovgaard:2001}) modified likelihood ratio statistics (\ref{skov}):
	\begin{equation}\label{gamma}
		\gamma(\psi_0) = \frac{2\{(\hat\sigma_{k0} - \hat\sigma_k)^\top \mathrm{Iss}(\hat\Omega_{0}^{-1})_{kk}^{-1}(\hat\sigma_{k0} - \hat\sigma_k)\}^{d/2}}{ \{-\log |\hat\Omega_k^{-1}\hat\Omega_0|\}^{d/2-1}    (\hat\omega_k - \hat\omega_{k0})^\top J_{kk}(\hat\sigma_{k0} - \hat\sigma_k)}
		\left\{\frac{|\mathrm{Iss}(\hat\Omega_{0}^{-1})_{kk}|}
		{|\mathrm{Iss}(\hat\Omega_{k}^{-1})_{kk}|}\right \}^{1/2}\,.
	\end{equation}

	\section{Computational aspects}\label{sec:com}\noindent
	\vspace{-1.5cm}
	\subsection{Calculation of the determinant of the Isserlis matrix}\label{sec:issdet}\noindent
	In situations where the dimension $p$ of the canonical parameter $\omega_k$  under the alternative model is smaller than $q^*$ but still { relatively} large, the calculation of the determinant of the matrix Iss$\{\hat\Omega_k^{-1}(t)\}_{kk}$ in (\ref{hfun}) can be {computationally quite demanding}. It is then advisable to exploit some useful results on the Isserlis matrix in order to speed up the computing time for the directional $p$-value.
	
	Let $A$ be a $q\times q$ symmetric invertible matrix. \citet[(15)]{rov98}, for any partition $(k', k'')$ of the edge set $k$ in (\ref{edgeset})
	such that $k' \cup k'' = k$ and $k' \cap k'' = \bar k$, show that
	\begin{equation*}
		|\mathrm{Iss}(A)_{kk}| = \dfrac{|\mathrm{Iss}(A)_{k'k'}||\mathrm{Iss}(A)_{k''k''}|}{|\mathrm{Iss}(A)_{\bar k\bar k}|}\,,
	\end{equation*}
	which gives a convenient way to reduce the dimensions of the matrices. If, moreover, the graph induced by  $k$ is chordal 
	with vertex set decomposable into cliques $C_1, \dots, C_{K}$ and separators $S_2, \dots,S_{K}$ according to definitions in \citet[Sect. 2.1]{Lauritzen:1996}, this can be further simplified to 
	\begin{equation} \label{decdet}
		|\mathrm{Iss}(A)_{kk}| = 2^{q} \dfrac{\prod_{i=1}^{K}|A_{C_i}|^{n_{C_i}+1}}{\prod_{i=2}^{K}|A_{S_i}|^{n_{S_i}+1}}\,,
	\end{equation}
	where $n_{C_i}$ and $n_{S_i}$ denote the number of nodes in the $i$th clique and $i$th separator, respectively, while
	$A_{C_i}$ and $A_{S_i}$ are submatrices of $A$ with rows and columns corresponding to the relative nodes
	\citep[(17)]{rov98}.
	
	\subsection{Numerical integration}
	\label{sec:tsup}\noindent
	{The upper bound $t_{\sup}$}  in (\ref{pval}) is the largest value of $t$ such that the ML estimate $\hat{\Omega}_k(t)$ 
	{ is positive definite}. By the same arguments as in \citet[Lemma 4.1]{huang21},
	this upper bound can be obtained explicitly as $t_{\sup}=1/(1-\nu_{(1)})$, where $\nu_{(1)}$ is the smallest of the $q$ eigenvalues 
	of $\hat{\Omega}_0 \hat{\Omega}_k^{-1}$.
	
	{Moreover, writing the integrand in (\ref{pval})  as $\exp \{\bar{g}(t;\psi)\}$, where $\bar{g}(t;\psi) =(d-1)\log t + \log h\{s(t);\psi\}$, we can improve the numerical stability of the calculations using} 
	the equivalent formula 
	\begin{eqnarray*}
		p(\psi)  = \frac{\int_{1}^{t_{\sup}} \exp \{\bar{g}(t;\psi) - \bar{g}(\hat{t};\psi)\} \text{d}t}{\int_{0}^{t_{\sup}} \exp \{\bar{g}(t;\psi) - \bar{g}(\hat{t};\psi)\} \text{d}t}\,, \qquad \text{where} \,\,\, \hat{t}=\underset{t \in [0, \ t_{\sup}]}{\arg\sup} \ \bar{g}(t;\psi)\,. 
	\end{eqnarray*}
	We have also found that the integrand function can be very concentrated around its mode, taking non-zero values in a shorter interval $[t_{\min},t_{\max}] \subseteq [0, t_{\sup}]$. To cope with this fact and deliver more stable numerical results,
	we  use Gauss--Hermite quadrature \citep{liu94} and integrate over $[t_{\min},t_{\max}]$ only. As a consequence, we compute the  
	directional $p$-value as
\begin{eqnarray}
	p(\psi) \;  \doteq  \; \frac{\int_{1}^{t_{\max}} \exp \{\bar{g}(t;\psi) - \bar{g}(\hat{t};\psi)\} \text{d}t}{\int_{t_{\min}}^{t_{\max}} \exp \{\bar{g}(t;\psi) - \bar{g}(\hat{t};\psi)\} \text{d}t}. 
	\label{pval:app}
\end{eqnarray}
{The choice $t_{\min} = \max\{0, \hat{t} - c/q(\hat{t};\psi)\}$ and $t_{\max} =  \min\{\hat{t} + c/q(\hat{t};\psi), t_{\sup}\}$, where $q({t};\psi) = - {\partial^2 \bar{g}(t;\psi)}/{\partial t^2} $ is reliable, with  $c$  a constant to be chosen  \citep[cf.][Sect. S1.3]{huang21}. The second derivative of the Isserlis determinant in the last factor of the integrand in (\ref{hfun})   cannot be derived explicitly
	and its numerical approximation may be unstable. In order to choose the width of the integration interval $[t_{\min},t_{\max}]$, we then set the function $q({t};\psi)$ equal only to the second derivative of the first factor in (\ref{hfun}), i.e.
	\begin{eqnarray*}
		q({t};\psi)=-\frac{\partial^2  |\hat\Omega^{-1}_k(t)|^{\frac{n-1}{2}}}{\partial t^2}   
		=\dfrac{d-1}{t^2}  +  \dfrac{n - 1}{2}\sum_{i=1}^q  \dfrac{(1-\nu_{i})^2}{(1-t+t\nu_{i})^{2}}\,.
	\end{eqnarray*}
	In our numerical experiments the value of $c$ was chosen for each pair $(n, q)$  by preliminary checks to ensure that integration from $t_{\min}$ to $t_{\max}$ was equal to that over  $[0, t_{\sup}]$, and then fixed for further simulations.  
	This simplification was found useful only in settings when $n>q$
	and cannot be applied  if $\bar g (t; \psi)$ is monotonic in $[0, t_{\sup}]$. The directional $p$-value in that case has to be calculated directly via formula (\ref{pval}), but this happened only 21 times in the Monte Carlo experiments below.

	\section{Simulation studies}\label{sec:sim}\noindent
	The performance of the directional approach in terms of covariance selection for Gaussian graphical models is examined here through  simulation-based experiments. 
	In the first scenario the focus is on a small chordal graph with $q=6$ nodes, similar to that in \citet[Ex.  7.3]{dawid93}. The two models under comparison, differing only by $d=3$ edges, are presented in Figure \ref{fig1}.  
	Monte Carlo simulations use $100\,000$ samples of size $n=8$ generated under the null hypothesis. The empirical $p$-value distribution of the tests based on $w(\psi_0)$, $w^*(\psi_0)$, $w^{**}(\psi_0)$ and the directional procedure is shown in the left plot of Figure \ref{fig2} with respect to the reference uniform distribution, zooming on the interval $(0, 0.1)$. The right plot compares the relative errors of the three most accurate methods. Despite the simplicity of the example, the likelihood ratio statistic leads to too many rejections of the null hypothesis because $n$ is relatively small. The higher-order modifications remedy this, yet the directional approach allows an exact control of the size of the test, up to numerical and Monte Carlo errors.

	\begin{figure}[t]
		\begin{center}
			\includegraphics[width=5.7cm]{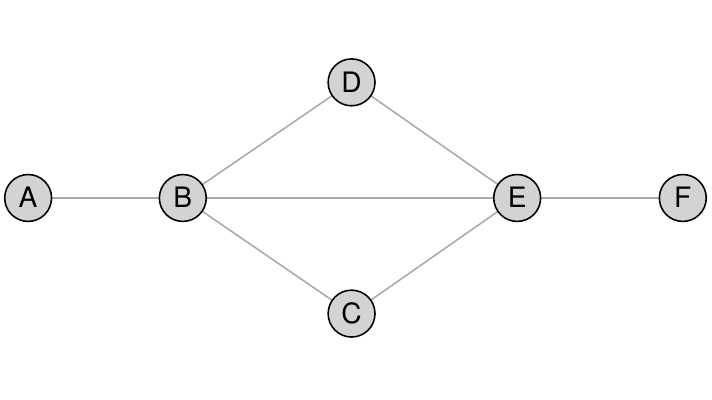} $\qquad\qquad$
			\includegraphics[width=5.7cm]{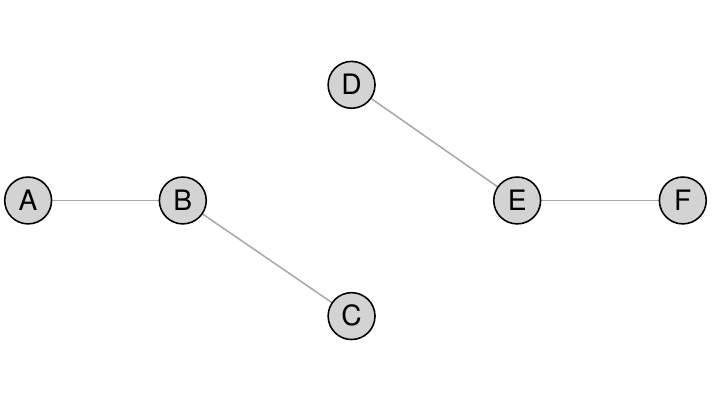}
			\vspace{-0.5cm}
			\caption{Graphs for the first simulation scenario where the dimension of the parameter of interest equals $d=3$. The alternative model for the chordal graph on the left is compared against the null model on the right.}
			\label{fig1}
		\end{center}
	\end{figure}
	
	\begin{figure}[t]
		\begin{center}
			\hspace{-1.2cm}
			\includegraphics[height=5.2cm]{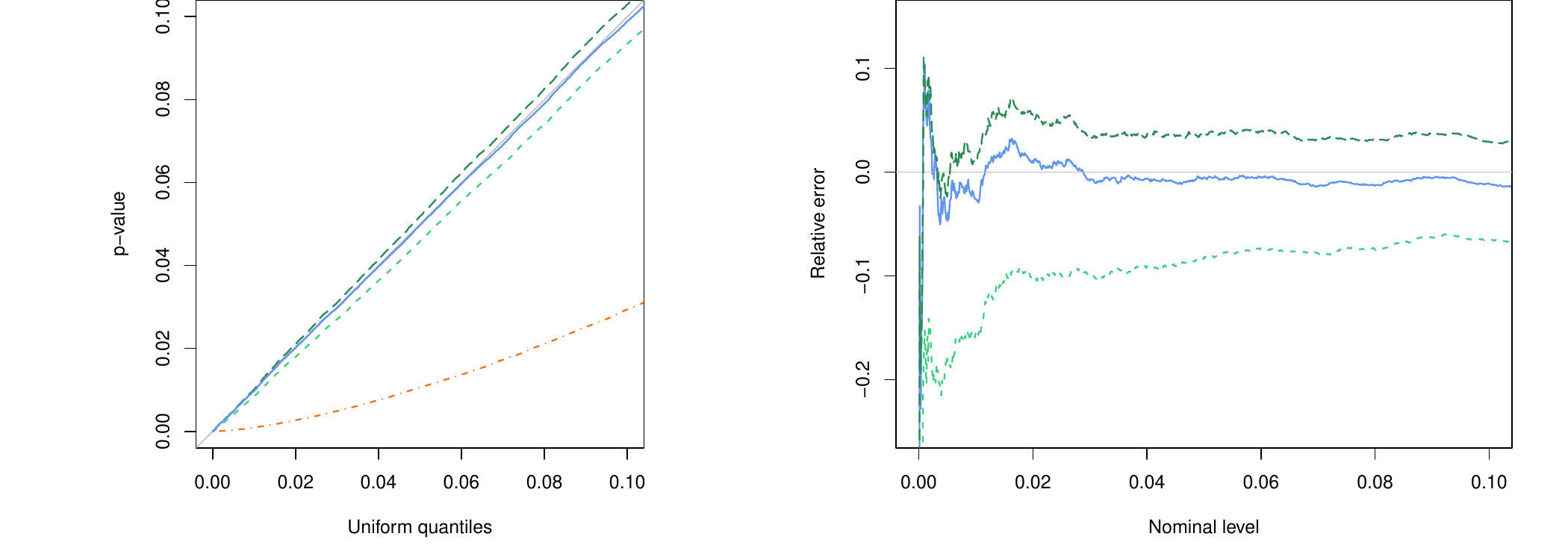}
			\vspace{-0.5cm}
			\caption{Results based on $100\,000$ samples
				simulated under the null model displayed on the right side of Figure \ref{fig1} with $n=8$ and $q=6$.
				On the left,  ordered empirical $p$-values $\hat p_{(i)}$ $(i = 1, \dots, 100\,000)$ smaller than 0.1 are compared with the uniform distribution on the diagonal for $w$ (red; dot-dashed), $w^*$ (green; dashed), $w^{**}$ (dark green; long-dashed) and the directional test (blue; solid). On the right, the corresponding relative errors 
				$\{\hat p_{({i)}} - (i/n)\}/(i/n)$
				are plotted in a similar fashion only for $w^*$, $w^{**}$  and the directional method.}
			\label{fig2}
		\end{center}
	\end{figure}
	
	\begin{table}[t]
		\caption{Empirical $p$-value distributions (\%) based on $100\,000$ replications. The Markovian model MD(1) is tested against different Markovian models of orders $m \in \{2,3,6,9\}$ under $H_1:\mathrm{MD}(m)$, when $n=60$ observations of a graph with $q=11$ nodes are available.}
		\label{tab1}
		\par
		\small
		\resizebox{\linewidth}{!}{\begin{tabular}{|lrrrrrrrrrrr|}
				\hline
				Nominal (\%) & 1.0 & 2.5 & 5.0 & 10.0 & 25.0 & 50.0 & 75.0 & 90.0 & 95.0 & 97.5 & 99.0 \\ 
				\hline
				vs MD(2), $d = 9$ &  &  & & & &  & &  &  & & \\ 
				Likelihood ratio, (\ref{lrtcovselect}) & 1.4 & 3.3 & 6.3 & 12.0 & 28.4 & 53.7 & 77.5 & 91.2 & 95.6 & 97.8 & 99.1 \\   
				Skovgaard's $w^*$, (\ref{gamma})& 1.0 & 2.5 & 5.1 & 10.0 & 25.1 & 50.2 & 75.1 & 89.9 & 94.9 & 97.4 & 99.0 \\   
				Skovgaard's $w^{**}$, (\ref{gamma}) & 1.0 & 2.5 & 5.1 & 10.0 & 25.1 & 50.2 & 75.1 & 89.9 & 94.9 & 97.4 & 99.0 \\   
				Directional, (\ref{pval:app})  & 1.0 & 2.5 & 5.1 & 10.0 & 25.2 & 50.3 & 75.2 & 90.1 & 95.0 & 97.5 & 99.0 \\   
				\hline
				vs MD(3), $d = 17$ &  &  & & & &  & &  &  & & \\ 
				Likelihood ratio, (\ref{lrtcovselect}) & 1.8 & 3.9 & 7.2 & 13.5 & 30.4 & 56.1 & 79.3 & 92.0 & 96.0 & 98.1 & 99.2 \\   
				Skovgaard's $w^*$, (\ref{gamma}) & 1.1 & 2.6 & 5.0 & 10.0 & 24.6 & 49.6 & 74.6 & 89.6 & 94.7 & 97.3 & 98.9 \\   
				Skovgaard's $w^{**}$, (\ref{gamma}) & 1.0 & 2.5 & 5.0 & 9.9 & 24.5 & 49.5 & 74.5 & 89.5 & 94.7 & 97.2 & 98.9 \\   
				Directional, (\ref{pval:app}) & 1.0 & 2.6 & 5.1 & 10.1 & 25.0 & 50.3 & 75.4 & 90.2 & 95.0 & 97.5 & 99.0 \\   
				\hline
				vs MD(6), $d = 35$ &  &  & & & &  & &  &  & & \\ 
				Likelihood ratio, (\ref{lrtcovselect}) & 2.5 & 5.5 & 9.8 & 17.4 & 36.2 & 62.2 & 83.3 & 94.0 & 97.2 & 98.6 & 99.5 \\   
				Skovgaard's $w^*$, (\ref{gamma}) & 0.8 & 2.1 & 4.3 & 8.8 & 22.4 & 46.4 & 71.7 & 87.8 & 93.6 & 96.6 & 98.5 \\   
				Skovgaard's $w^{**}$, (\ref{gamma}) & 0.8 & 2.1 & 4.2 & 8.6 & 22.0 & 45.9 & 71.2 & 87.5 & 93.4 & 96.4 & 98.5 \\   
				Directional, (\ref{pval:app}) & 1.0 & 2.5 & 4.9 & 10.0 & 25.0 & 50.3 & 75.3 & 90.2 & 95.1 & 97.5 & 99.0 \\   
				\hline
				vs MD(9), $d = 44$ &  &  & & & &  & &  &  & & \\ 
				Likelihood ratio, (\ref{lrtcovselect}) & 3.3 & 6.9 & 12.0 & 20.6 & 40.8 & 66.2 & 85.9 & 95.2 & 97.8 & 99.0 & 99.6 \\   
				Skovgaard's $w^*$, (\ref{gamma}) & 0.7 & 1.8 & 3.7 & 7.8 & 20.7 & 43.7 & 69.1 & 86.3 & 92.6 & 96.1 & 98.2 \\   
				Skovgaard's $w^{**}$, (\ref{gamma}) & 0.7 & 1.8 & 3.6 & 7.5 & 20.1 & 42.8 & 68.2 & 85.7 & 92.2 & 95.8 & 98.1 \\   
				Directional, (\ref{pval:app})  & 1.0 & 2.4 & 4.9 & 9.9 & 25.2 & 50.0 & 75.0 & 90.1 & 95.1 & 97.5 & 99.0 \\
				\hline
				Standard error& 0.0 & 0.0 & 0.1 & 0.1 & 0.1 & 0.2 & 0.1 & 0.1 & 0.1 & 0.0 & 0.0 \\
				\hline
		\end{tabular}}
	\end{table}
	
	The inferential benefits of our proposal over the omnibus likelihood-based competitors are particularly appreciated with high magnitudes of $q$ and $d$.
	The second  scenario is based on the data of \citet[Tab. 1]{Kenward87} from a study on intestinal parasites of 60 calves, where the weight in kg of each bovine was recorded on 11 occasions during the grazing season.
	To enable comparison with \citet[Sect.  5.3]{dav14}, who could only test the saturated model,
	we draw $100\,000$ samples of size $n=60$
	from a $q$-variate Gaussian random variable
	under the hypothesis of first-order Markovian dependence MD(1) with tridiagonal concentration matrix. For each $q\in\{11, 30, 50\}$, the null hypothesis $H_0: \mathrm{MD}(1)$ 
	is tested against four different alternative unsaturated structures, using also $w(\psi_0)$, $w^*(\psi_0)$ and $w^{**}(\psi_0)$. These Markovian dependence models of order $m$  under $H_1: \mathrm{MD}(m)$ with $1<m<q-1$ correspond to so-called band concentration matrices, whose nonzero entries are confined to $m$ diagonals on either side of the main one. The orders $m$ are chosen to check the behavior of the various methods for a wide range of dimensions $d$ of the parameter of interest, and consequently of the nuisance component. 
	Since the Markovian structure induces a chordal graph,
	the simplification (\ref{decdet}) is particularly useful for computing the directional $p$-values with such a high-dimensional parameter of interest.

	Table \ref{tab1} reports experimental results obtained when $q=11$ as in the original dataset, whereas Tables \ref{tab2} and \ref{tab3} refers to cases with data simulated using a larger covariance matrix, $q=30$ and $q=50$ respectively. In line with 
	our theoretical findings,
	the empirical distribution of the directional $p$-values is  essentially uniform in all settings, and almost unaffected by the size of $q$ and $d$. 
	The usual likelihood ratio statistic $w(\psi_0)$ is very sensitive to the dimension of both $\psi$ and $\lambda$; its adjustments $w^*(\psi_0)$ and, particularly, $w^{**}(\psi_0)$ seem to suffer from the increasing dimension $d$ of the parameter of interest. Tables \ref{tab2} and \ref{tab3} clearly indicate that, as $d$ grows, the test based on $w(\psi_0)$ becomes too liberal and those based on $w^*(\psi_0)$ and $w^{**}(\psi_0)$ too conservative. 
	For the intermediate case $q=30$, the leftmost panels of Figure \ref{fig3} contrasts the null empirical distribution of the directional $p$-values with those from $w(\psi_0)$, $w^*(\psi_0)$ and $w^{**}(\psi_0)$.  The almost perfect agreement of our proposal with the benchmark uniform distribution given by the diagonal of the panels is apparent.
	
	\begin{table}[t]
		\caption{Empirical $p$-value distributions (\%) based on $100\,000$ replications.  The first-order Markovian model under $H_0: \mathrm{MD}(1)$ is tested against different Markovian models of orders $m \in \{2,9,18,28\}$ under $H_1:\mathrm{MD}(m)$, when $n=60$ observations of a graph with $q=30$ nodes are available.}
		\label{tab2}
		\par
		\resizebox{\linewidth}{!}{\begin{tabular}{|lrrrrrrrrrrr|}
				\hline
				Nominal (\%) & 1.0 & 2.5 & 5.0 & 10.0 & 25.0 & 50.0 & 75.0 & 90.0 & 95.0 & 97.5 & 99.0 \\ 
				\hline
				vs MD(2), $d = 28$ &  &  & & & &  & &  &  & & \\ 
				Likelihood ratio, (\ref{lrtcovselect}) & 1.6 & 3.8 & 7.2 & 13.4 & 30.5 & 56.4 & 79.4 & 92.2 & 96.2 & 98.1 & 99.3 \\   
				Skovgaard's $w^*$, (\ref{gamma}) & 1.0 & 2.5 & 5.0 & 10.0 & 24.9 & 50.0 & 75.1 & 90.1 & 95.0 & 97.5 & 99.0 \\   
				Skovgaard's $w^{**}$, (\ref{gamma}) & 1.0 & 2.5 & 5.0 & 10.0 & 24.9 & 50.0 & 75.0 & 90.0 & 95.0 & 97.5 & 99.0 \\   
				Directional, (\ref{pval:app}) & 1.0 & 2.4 & 4.9 & 10.0 & 24.9 & 50.1 & 75.2 & 90.2 & 95.1 & 97.5 & 99.0 \\ 
				\hline
				vs MD(9), $d = 196$ &  &  & & & &  & &  &  & & \\ 
				Likelihood ratio, (\ref{lrtcovselect})   & 11.1 & 19.1 & 28.4 & 41.5 & 64.6 & 84.6 & 95.3 & 98.7 & 99.5 & 99.8 & 99.9 \\   
				Skovgaard's $w^*$, (\ref{gamma}) & 0.3 & 0.9 & 2.0 & 4.4 & 13.3 & 32.3 & 57.9 & 78.5 & 87.1 & 92.5 & 96.4 \\   
				Skovgaard's $w^{**}$, (\ref{gamma}) & 0.3 & 0.8 & 1.7 & 3.9 & 12.1 & 30.2 & 55.4 & 76.5 & 85.7 & 91.4 & 95.7 \\   
				Directional, (\ref{pval:app})  & 0.9 & 2.3 & 4.8 & 9.7 & 24.7 & 50.3 & 75.8 & 90.5 & 95.4 & 97.7 & 99.1 \\
				\hline
				vs MD(18), $d = 340$ &  &  & & & &  & &  &  & & \\ 
				Likelihood ratio, (\ref{lrtcovselect})   &  53.8 & 66.9 & 76.9 & 86.0 & 95.0 & 98.8 & 99.8 & 100.0 & 100.0 & 100.0 & 100.0 \\   
				Skovgaard's $w^*$, (\ref{gamma}) & 0.0 & 0.1 & 0.3 & 0.7 & 3.0 & 10.7 & 27.4 & 48.8 & 62.1 & 72.8 & 82.9 \\   
				Skovgaard's $w^{**}$, (\ref{gamma}) & 0.0 & 0.0 & 0.1 & 0.4 & 1.7 & 6.9 & 19.5 & 38.2 & 51.2 & 62.5 & 74.3 \\   
				Directional, (\ref{pval:app})  & 0.8 & 2.2 & 4.6 & 9.5 & 24.7 & 50.2 & 76.0 & 90.8 & 95.6 & 97.8 & 99.2 \\
				\hline
				vs MD(28), $d = 405$ &  &  & & & &  & &  &  & & \\ 
				Likelihood ratio, (\ref{lrtcovselect})  & 86.2 & 92.3 & 95.6 & 97.9 & 99.5 & 99.9 & 100.0 & 100.0 & 100.0 & 100.0 & 100.0 \\   
				Skovgaard's $w^*$, (\ref{gamma}) & 0.0 & 0.0 & 0.0 & 0.2 & 0.9 & 4.3 & 13.8 & 30.0 & 42.5 & 53.9 & 67.0 \\   
				Skovgaard's $w^{**}$, (\ref{gamma})& 0.0 & 0.0 & 0.0 & 0.0 & 0.2 & 1.4 & 5.9 & 15.5 & 24.5 & 33.9 & 46.4 \\   
				Directional, (\ref{pval:app})  & 1.0 & 2.4 & 5.1 & 10.1 & 25.2 & 50.1 & 75.1 & 90.1 & 95.1 & 97.5 & 99.0 \\
				\hline
				Standard error& 0.0 & 0.0 & 0.1 & 0.1 & 0.1 & 0.2 & 0.1 & 0.1 & 0.1 & 0.0 & 0.0 \\
				\hline
		\end{tabular}}
	\end{table}
	
	Before proceeding, let us focus on the implementation of formula (\ref{decdet}) to obtain the determinant of the Isserlis matrix of $\Omega_k^{-1}$ estimated under the alternative hypothesis. When multiplying the determinants of many square matrices of moderate order, some propagation of numerical errors can occur. In our experiments this is visible, to a certain extent, in the intermediate sections of Tables \ref{tab2} and \ref{tab3}, when the performance of directional tests seems slightly less excellent than in the remaining sections. Indeed, when the null is tested against more extreme Markovian models, the matrices involved in (\ref{decdet}) are either many but small (top section) or large but few (bottom section), thus the final product of their determinants is not overly affected by numerical error. That being said, it is important to point out that in all settings the directional approach remains remarkably accurate and brings a great improvement over the competing testing procedures. 
	
	The third simulation scenario considers a block diagonal configuration of the concentration matrix under the null hypothesis. Here, $100\,000$ samples of size $n \in \{40, 60, 90, 120\}$ were drawn from a 
	normal distribution with $q=50$ components
	and covariance matrix $\Sigma_{0}=\diag\{\Sigma_{01},\Sigma_{01}\}$, with $\Sigma_{01}$ sub-matrix $25\times 25$ having diagonal entries equal to 1 and off-diagonal entries equal to 0.5.
	Such condition clearly implies that $\Omega_0=\Sigma^{-1}_{0}$ is also block diagonal, so that the first 25 nodes are conditionally (as well as unconditionally) independent of the last 25 nodes in the graph.
	On the other hand, our alternative model admits the existence of some conditional dependence between the two subsets of nodes. Specifically, besides the nonzero elements defined in $\Omega_0$, we also suppose $\Omega_{ij}=\Omega_{ji} \neq 0$ for $i=16,\dots,25$ and $j=26,\dots,50$. It follows that the dimension of the parameter of interest is $d=250$ and  (\ref{decdet}) can be used to speed up calculations of the Isserlis matrix associated with the chordal alternative incomplete graph.
	
	Simulation results in this framework are presented in Table \ref{tab4}. Given the notable size of $d$, the relative performance of the approximations under comparison, in terms of the empirical $p$-value distribution, is analogous to that in the previous experiment, with the only exception that here the version $w^{**}(\psi_0)$ appears generally more reliable than $w^{*}(\psi_0)$. Although the increase in sample size generates some accuracy improvements for all the competitors as expected, the  empirical directional $p$-value guarantees an almost perfect agreement with  its theoretical uniform distribution for all values of $n$ considered. The extreme liberality of the standard likelihood ratio test persists, Skovgaard's $w^{*}(\psi_0)$ does not correct it enough and the version $w^{**}(\psi_0)$ overcorrects it.
	Like before, the rightmost panels of Figure \ref{fig3} displays the $p$-values obtained via the likelihood ratio statistic, its modified versions and the directional procedure.

	\begin{figure}[t]
		\begin{center}
			\includegraphics[height=8.6cm]{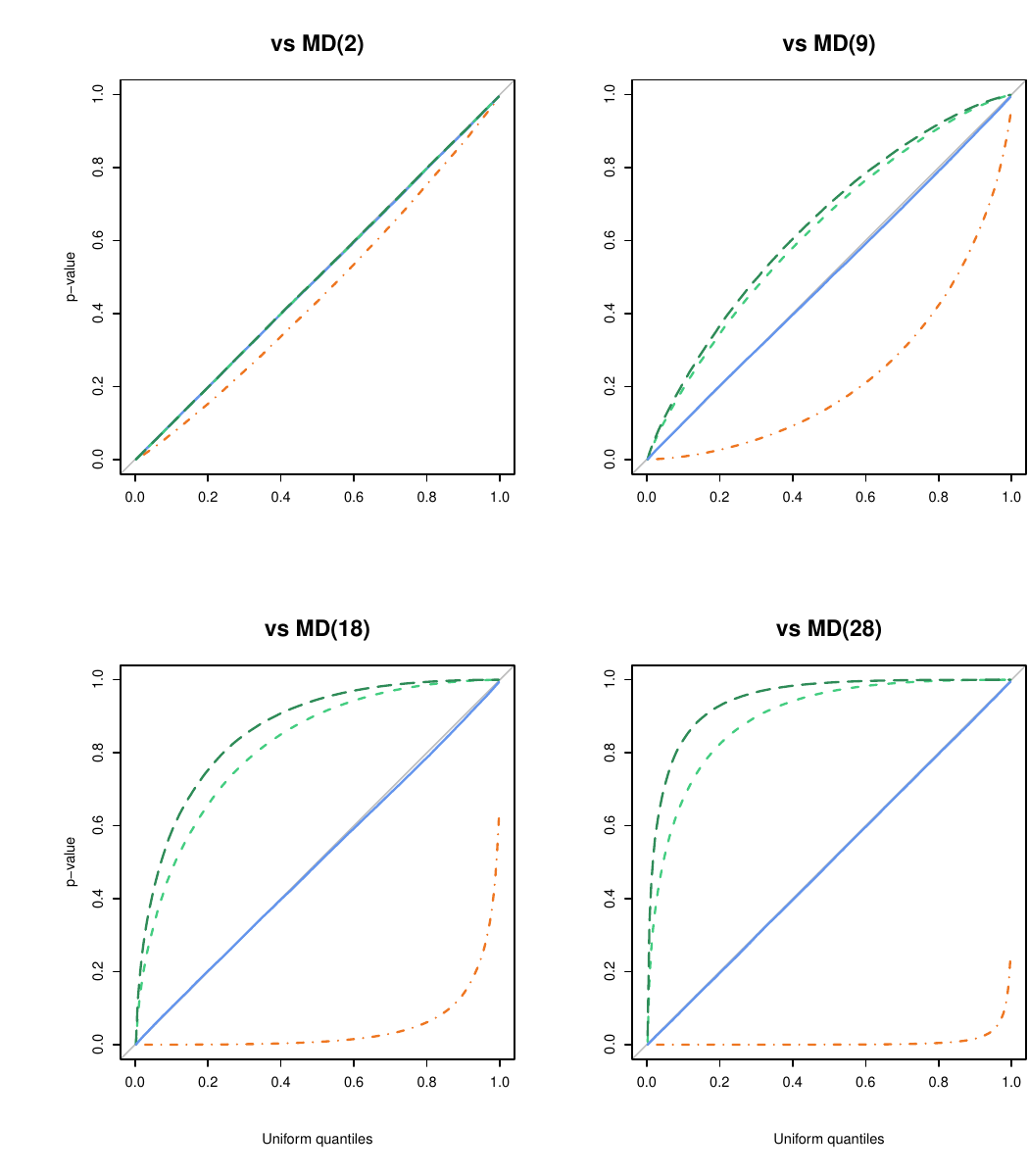}
			\hspace{0.5cm}
			\includegraphics[height=8.6cm]{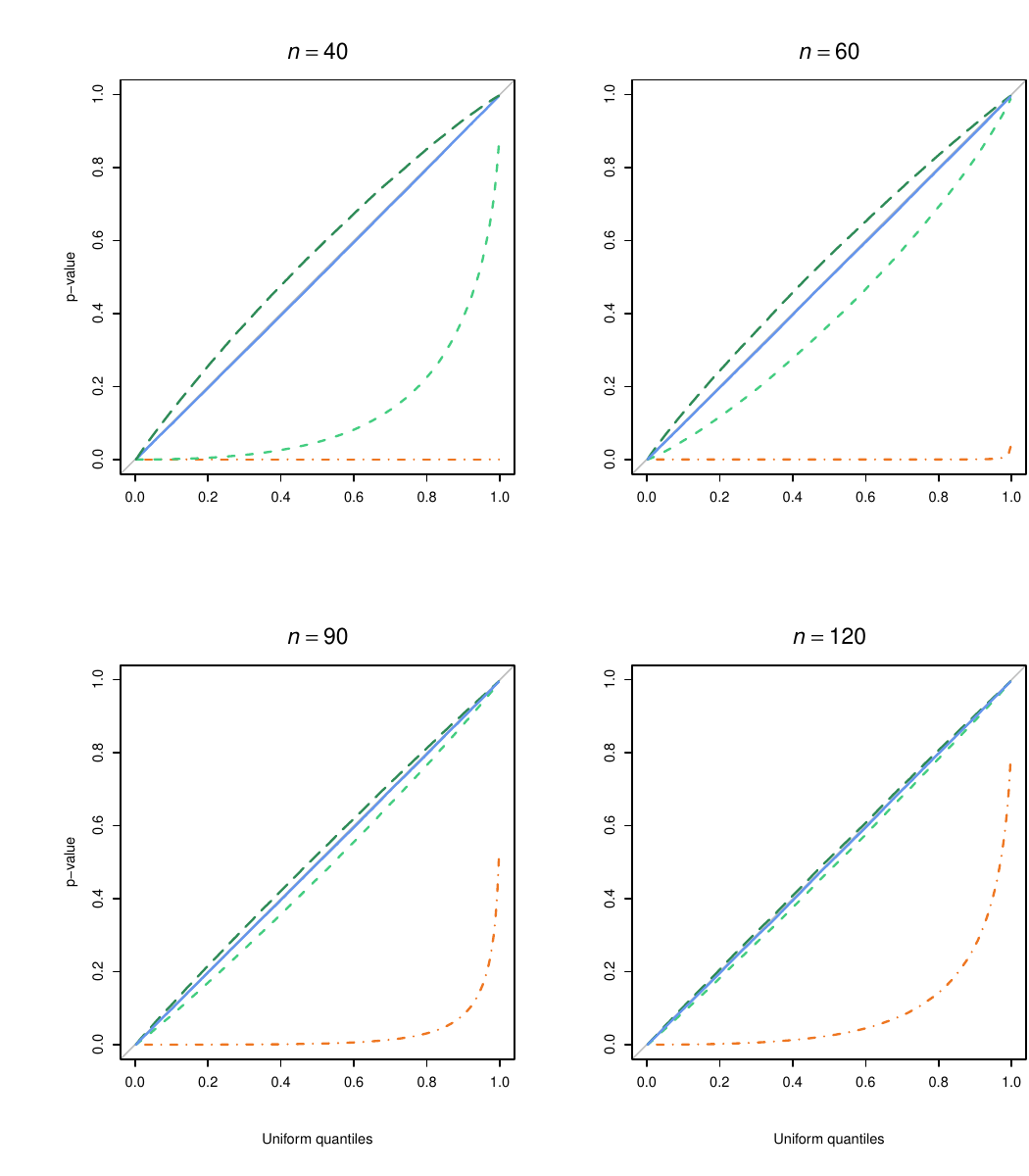}
			\vspace{-0.5cm}
			\caption{Results based on $100\,000$ simulated samples. In all eight panels, the empirical $p$-values obtained via $w$ (dot-dashed), $w^*$ (dashed), $w^{**}$ (long-dashed) and the directional test (solid) are compared with the uniform distribution given by the diagonal.
				Leftmost panels: the null model MD(1) assumes first-order Markovian dependence with $n=60$ and $q=30$. The four panels correspond to different Markovian models under the alternative hypothesis and related dimensions of $\psi$: MD(2) and $d=28$ (top left), MD(9) and $d=196$ (top right), MD(18) and $d=340$ (bottom left), MD(28) and $d=405$ (bottom right).
				Rightmost panels: the null model assuming a block diagonal concentration matrix with $q=50$ is tested against the same alternative hypothesis implying $d=250$. The four panels correspond to
				different sample sizes: $n=40$ (top left), $n=60$ (top right),
				$n=90$ (bottom left), $n=120$ (bottom right).}
			\label{fig3}
		\end{center}
	\end{figure}

	\begin{table}[t]
		\caption{ 
			Empirical $p$-value distributions (\%) based on $100\,000$ replications. The first-order Markovian model under $H_0: \mathrm{MD}(1)$ is tested against different Markovian models of orders $m \in \{2,16,32,48\}$ under $H_1:\mathrm{MD}(m)$, when $n=60$ observations of a graph with $q=50$ nodes are available.}
		\label{tab3}
		\par
		\resizebox{\linewidth}{!}{\begin{tabular}{|lrrrrrrrrrrr|}
				\hline
				Nominal (\%) & 1.0 & 2.5 & 5.0 & 10.0 & 25.0 & 50.0 & 75.0 & 90.0 & 95.0 & 97.5 & 99.0 \\ 
				\hline
				vs MD(2), $d = 48$ &  &  & & & &  & &  &  & & \\ 
				Likelihood ratio, (\ref{lrtcovselect}) & 1.8 & 4.2 & 7.8 & 14.5 & 32.4 & 58.2 & 80.9 & 93.0 & 96.7 & 98.4 & 99.4 \\   
				Skovgaard's $w^*$, (\ref{gamma}) & 1.0 & 2.5 & 5.0 & 9.9 & 25.1 & 50.1 & 74.9 & 90.0 & 95.1 & 97.5 & 99.0 \\   
				Skovgaard's $w^{**}$, (\ref{gamma})  & 1.0 & 2.5 & 5.0 & 9.9 & 25.0 & 50.0 & 74.9 & 89.9 & 95.1 & 97.5 & 99.0 \\   
				Directional, (\ref{pval:app})  & 1.0 & 2.5 & 4.9 & 9.9 & 25.0 & 50.1 & 75.1 & 90.1 & 95.2 & 97.6 & 99.1 \\
				\hline
				vs MD(16), $d = 615$ &  &  & & & &  & &  &  & & \\ 
				Likelihood ratio, (\ref{lrtcovselect})   & 77.9 & 86.7 & 92.1 & 96.0 & 99.0 & 99.8 & 100.0 & 100.0 & 100.0 & 100.0 & 100.0 \\   
				Skovgaard's $w^*$, (\ref{gamma})  & 0.0 & 0.0 & 0.1 & 0.2 & 1.1 & 5.1 & 15.9 & 33.3 & 46.2 & 57.9 & 70.6 \\   
				Skovgaard's $w^{**}$, (\ref{gamma}) & 0.0 & 0.0 & 0.0 & 0.1 & 0.5 & 2.6 & 9.3 & 22.2 & 33.2 & 44.2 & 57.4 \\   
				Directional, (\ref{pval:app})  & 0.8 & 2.0 & 4.3 & 9.1 & 24.4 & 50.4 & 76.3 & 91.4 & 96.1 & 98.1 & 99.3 \\
				\hline
				vs MD(32), $d = 1023$ &  &  & & & &  & &  &  & & \\ 
				Likelihood ratio, (\ref{lrtcovselect})   & 100.0 & 100.0 & 100.0 & 100.0 & 100.0 & 100.0 & 100.0 & 100.0 & 100.0 & 100.0 & 100.0 \\   
				Skovgaard's $w^*$, (\ref{gamma})  & 0.0 & 0.0 & 0.0 & 0.0 & 0.0 & 0.0 & 0.2 & 0.7 & 1.7 & 3.4 & 6.7 \\   
				Skovgaard's $w^{**}$, (\ref{gamma}) & 0.0 & 0.0 & 0.0 & 0.0 & 0.0 & 0.0 & 0.0 & 0.0 & 0.0 & 0.1 & 0.2 \\   
				Directional, (\ref{pval:app}) & 0.5 & 1.4 & 3.4 & 8.0 & 23.5 & 51.7 & 78.6 & 92.8 & 96.9 & 98.7 & 99.5 \\ 
				\hline
				vs MD(48), $d = 1175$ &  &  & & & &  & &  &  & & \\ 
				Likelihood ratio, (\ref{lrtcovselect})  & 100.0 & 100.0 & 100.0 & 100.0 & 100.0 & 100.0 & 100.0 & 100.0 & 100.0 & 100.0 & 100.0 \\   
				Skovgaard's $w^*$, (\ref{gamma})  & 0.0 & 0.0 & 0.0 & 0.0 & 0.0 & 0.0 & 0.1 & 0.4 & 1.0 & 2.0 & 4.0 \\   
				Skovgaard's $w^{**}$, (\ref{gamma}) & 0.0 & 0.0 & 0.0 & 0.0 & 0.0 & 0.0 & 0.0 & 0.0 & 0.0 & 0.0 & 0.0 \\   
				Directional, (\ref{pval:app}) & 0.8 & 2.2 & 4.7 & 9.8 & 25.4 & 51.1 & 76.2 & 90.9 & 95.5 & 97.8 & 99.2 \\ 
				\hline
				Standard error& 0.0 & 0.0 & 0.1 & 0.1 & 0.1 & 0.2 & 0.1 & 0.1 & 0.1 & 0.0 & 0.0 \\
				\hline
		\end{tabular}}
	\end{table}
	
	\begin{table}[t]
		\caption{Empirical $p$-value distributions (\%) based on $100\,000$ replications. The two-block diagonal structure of the concentration matrix for a graph with $q=50$ nodes is tested against a more complex structure including $d=250$ additional edges.}
		\label{tab4}
		\par
		\resizebox{\linewidth}{!}{\begin{tabular}{|lrrrrrrrrrrr|}
				\hline
				Nominal (\%) & 1.0 & 2.5 & 5.0 & 10.0 & 25.0 & 50.0 & 75.0 & 90.0 & 95.0 & 97.5 & 99.0 \\ 
				\hline
				$n = 40$ 
				&  &  & & & &  & &  &  & & \\ 
				Likelihood ratio, (\ref{lrtcovselect}) & 100.0 & 100.0 & 100.0 & 100.0 & 100.0 & 100.0 & 100.0 & 100.0 & 100.0 & 100.0 & 100.0 \\   
				Skovgaard's $w^{*}$, (\ref{gamma}) & 27.1 & 39.1 & 50.6 & 63.5 & 81.7 & 93.7 & 98.5 & 99.7 & 99.9 & 100.0 & 100.0 \\   
				Skovgaard's $w^{**}$, (\ref{gamma}) & 0.7 & 1.7 & 3.4 & 7.2 & 19.3 & 42.0 & 68.0 & 85.9 & 92.5 & 96.1 & 98.3 \\   
				Directional, (\ref{pval:app}) & 1.0 & 2.5 & 5.0 & 10.1 & 25.2 & 50.2 & 75.2 & 90.0 & 94.9 & 97.4 & 98.9 \\
				\hline
				$n = 60$ 
				&  &  & & & &  & &  &  & & \\ 
				Likelihood ratio, (\ref{lrtcovselect})  & 98.4 & 99.3 & 99.7 & 99.9 & 100.0 & 100.0 & 100.0 & 100.0 & 100.0 & 100.0 & 100.0 \\   
				Skovgaard's $w^{*}$, (\ref{gamma}) & 2.4 & 5.3 & 9.6 & 17.3 & 36.6 & 62.9 & 84.2 & 94.6 & 97.6 & 98.9 & 99.6 \\   
				Skovgaard's $w^{**}$, (\ref{gamma}) & 0.6 & 1.7 & 3.5 & 7.5 & 20.4 & 43.9 & 70.3 & 87.4 & 93.5 & 96.7 & 98.6 \\   
				Directional, (\ref{pval:app}) & 1.0 & 2.5 & 5.0 & 10.0 & 25.1 & 50.1 & 75.2 & 90.2 & 95.1 & 97.6 & 99.0 \\
				\hline
				$n = 90$ 
				&  &  & & & &  & &  &  & & \\
				Likelihood ratio, (\ref{lrtcovselect})  & 65.9 & 77.1 & 85.0 & 91.5 & 97.3 & 99.4 & 99.9 & 100.0 & 100.0 & 100.0 & 100.0 \\   
				Skovgaard's $w^{*}$, (\ref{gamma}) & 1.3 & 3.2 & 6.1 & 12.0 & 28.5 & 54.2 & 78.2 & 91.7 & 96.0 & 98.1 & 99.2 \\   
				Skovgaard's $w^{**}$, (\ref{gamma}) & 0.8 & 2.1 & 4.3 & 8.9 & 23.0 & 47.6 & 73.2 & 89.0 & 94.5 & 97.2 & 98.8 \\   
				Directional, (\ref{pval:app}) & 0.9 & 2.5 & 5.0 & 10.1 & 25.0 & 50.1 & 75.1 & 90.1 & 95.1 & 97.6 & 99.0 \\
				\hline
				$n = 120$ 
				&  &  & & & &  & &  &  & & \\
				Likelihood ratio, (\ref{lrtcovselect}) & 36.6 & 50.0 & 61.6 & 73.6 & 88.6 & 96.7 & 99.3 & 99.9 & 100.0 & 100.0 & 100.0 \\   
				Skovgaard's $w^{*}$, (\ref{gamma}) & 1.1 & 2.9 & 5.6 & 11.0 & 26.8 & 52.2 & 76.5 & 90.9 & 95.5 & 97.8 & 99.1 \\   
				Skovgaard's $w^{**}$, (\ref{gamma}) & 0.9 & 2.3 & 4.6 & 9.4 & 24.0 & 48.7 & 73.9 & 89.4 & 94.6 & 97.3 & 98.9 \\   
				Directional, (\ref{pval:app}) & 1.0 & 2.5 & 5.0 & 10.1 & 25.1 & 50.1 & 75.0 & 90.0 & 95.0 & 97.5 & 99.0 \\
				\hline
				Standard error& 0.0 & 0.0 & 0.1 & 0.1 & 0.1 & 0.2 & 0.1 & 0.1 & 0.1 & 0.0 & 0.0 \\
				\hline
		\end{tabular}}
	\end{table}
	
	As an empirical check of the accuracy of our proposal for non-decomposable models, we consider in
	the fourth simulation scenario  a small non-chordal graph with $q=4$ nodes as in \citet[Sect.  4]{erik96}. Figure \ref{fig5} displays the two models under comparison, which differ only by $d=2$ edges. 
	Setting the sample size to $n=7$, $100\,000$ artificial samples are simulated
	under the null hypothesis. As for the first scenario, results are presented in two panels of Figure \ref{fig6}.
	Since $n$ is small with respect to $q$ and $d$,  the chi-squared approximation to the distribution of the likelihood ratio statistic is misleading. The improved versions of $w$, especially $w^*$ here, are more reliable. However, even in this application to a non-chordal graph, the superiority of the directional approach based on the accurate saddlepoint approximation is evident in terms of relative error. 
	

	\begin{figure}[t]
		\begin{center}
			\includegraphics[height=4cm]{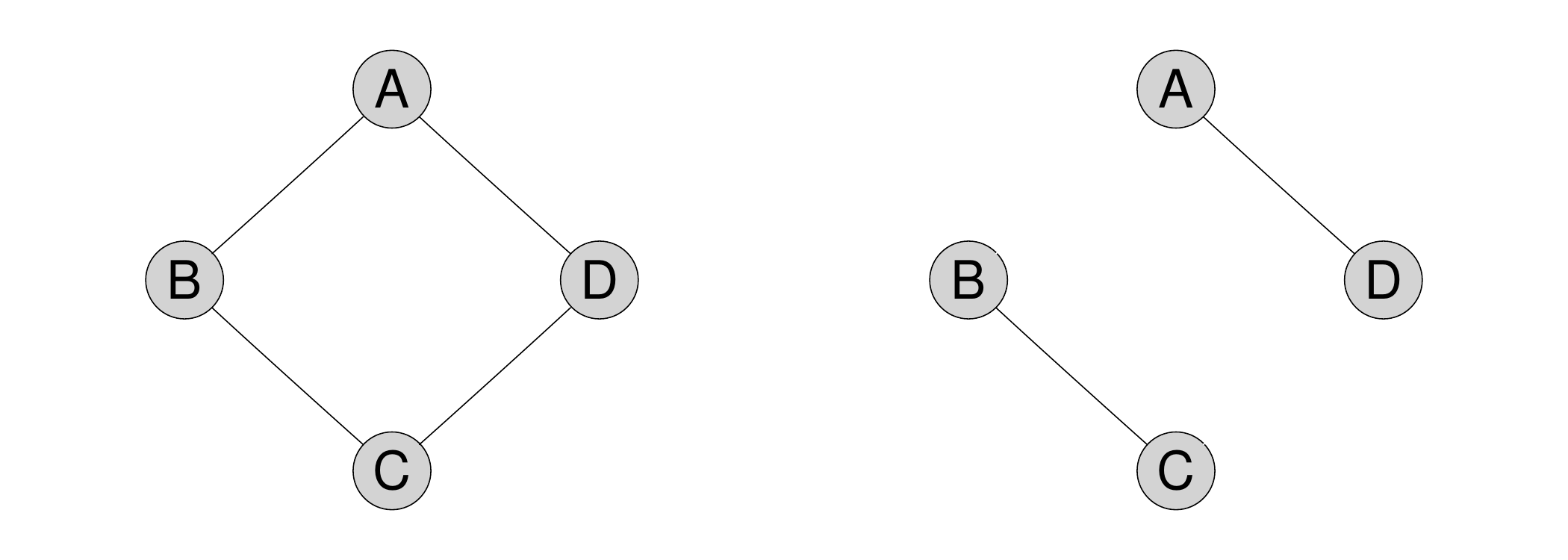}
			\vspace{-0.5cm}
			\caption{Graphs for the fourth simulation scenario where the dimension of the parameter of interest equals $d=2$. The alternative model for the non-chordal graph on the left is compared against the null model on the right.}
			\label{fig5}
		\end{center}
	\end{figure}
	
	\begin{figure}[t]
		\begin{center}
			\hspace{-1.2cm}
			\includegraphics[height=5.2cm]{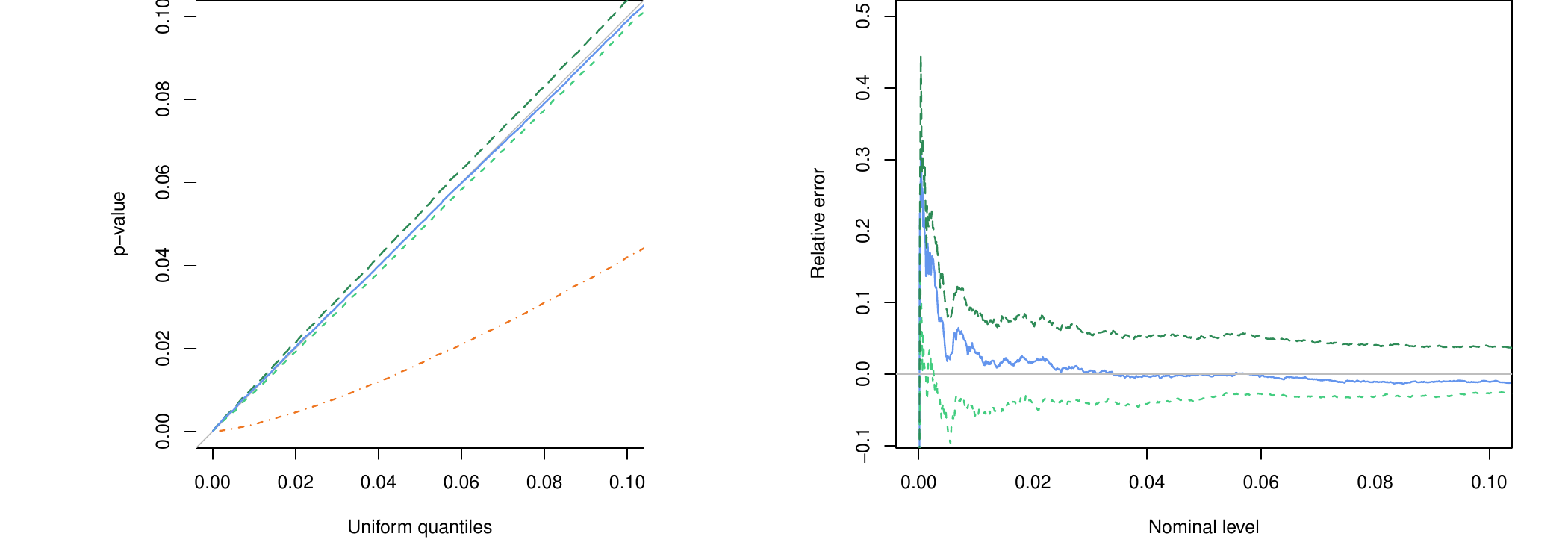}
			\vspace{-0.5cm}
			\caption{Results based on $100\,000$ samples
				simulated under the null model represented by the right graph of Figure \ref{fig5} with $n=7$ and $q=4$.
				On the left, ordered empirical $p$-values $\hat p_{(i)}$ $(i = 1, \dots, 100\,000)$ smaller than 0.1 are compared with the uniform distribution on the diagonal for $w$ (red; dot-dashed), $w^*$ (green; dashed), $w^{**}$ (dark green; long-dashed) and the directional test (blue; solid). On the right, the corresponding relative errors 
				$\{\hat p_{({i)}} - (i/n)\}/(i/n)$
				are plotted in a similar fashion only for $w^*$, $w^{**}$  and the directional method. } 
		\label{fig6}
	\end{center}
\end{figure}

\section{Applications}\label{sec:app}
First, we examine the dataset already introduced in the second simulation scenario of Section \ref{sec:sim} from the experiment about the control of intestinal parasites in cattle \citep[Tab. 1]{Kenward87}. However, here we focus on the two treatment groups with equal size $n=30$ separately, in order to investigate differences in the underlying temporal dynamics of growth. Recalling that each animal was weighed $q=11$ consecutive times, we start by assuming a Markovian dependence of order $m=3$, the simplest model accepted in a test against the saturated one
by all the procedures under analysis and in both groups.
This model is then compared against the null hypothesis of first-order dependence, implying $d=17$. For the calves randomly assigned to the first treatment, the likelihood ratio statistic is $w(\psi_0)=28.384$ with $p$-value$=0.041$, Skovgaard's modifications are $w^*(\psi_0)=22.977$ with $p$-value$=0.150$ and $w^{**}(\psi_0)=22.691$ with $p$-value$=0.160$ and the directional $p$-value is $0.111$.
For the second group we get instead $w(\psi_0)=31.895$ with $p$-value$=0.016$, $w^*(\psi_0)=30.055$ with $p$-value$=0.026$, $w^{**}(\psi_0)=30.028$ with $p$-value$=0.026$ and directional $p$-value$=0.029$.
The standard likelihood ratio test is the only one to reject the MD(1) model at a 5\% significance level for both treatments. Conversely, the other statistics recognize a different time pattern and indicate a more complex dependence of the weights in the second group. 

We now consider
some microarray data from the biostatistical literature \citep[see, e.g.,][]{massa10}, which characterize gene expression signatures in acute lymphocytic leukemia 
cells associated with genotypic abnormalities in adult patients. The normalized version of such data, available in the package topologyGSA \citep{top16} of the R software \citep{R}, is especially useful for analyzing the B-cell receptor (BCR) signaling pathway composed by $q=35$ gene products. The observed samples are classified according to the presence of molecular rearrangements in their genetic profile. 

The conversion of biological pathways into graphical models has become standard practice in biostatistics to separate and compare specific portions of the genetic process under examination. Based on findings in \citet{massa10}, it seems of interest to investigate whether the graph resulting from the well-known BCR signaling pathway in Figure \ref{fig7} can be further simplified. In more detail,
the restricted graphical model under the null hypothesis in our analysis corresponds to
the identified path starting from nodes CD22 and CD72 and ending at AP1, going through RasGRP3, Ras, Raf, MEK1/2 and ERK enzymes. Such a comparison implies testing the lack  of $d=12$ edges and can be carried out 
on the subset of patients not suffering from so-called BCR/ABL rearrangements. 
With
$n=41$, we obtain $w(\psi_0)=33.520$ with $p$-value$=8.028 \times 10^{-4}$, $w^*(\psi_0)=32.172$ with $p$-value$=13.018  \times 10^{-4}$,
$w^{**}(\psi_0)=32.158$ with $p$-value$=13.083  \times 10^{-4}$
and directional $p$-value$=13.941 \times 10^{-4}$.
%
Although all four methods indicate that the data are not consistent with the shorter biological path, the $p$-value from usual likelihood ratio test $w(\psi_0)$ is relatively much smaller than the other three, and in these types of problems very small $p$-values are relevant. The agreement of Skovgaard's approximations with the directional $p$-value is consistent with our simulations results for  small values of $d$ with respect to $n$.

\begin{figure}[t]
	\begin{center}
		\includegraphics[height=15cm]{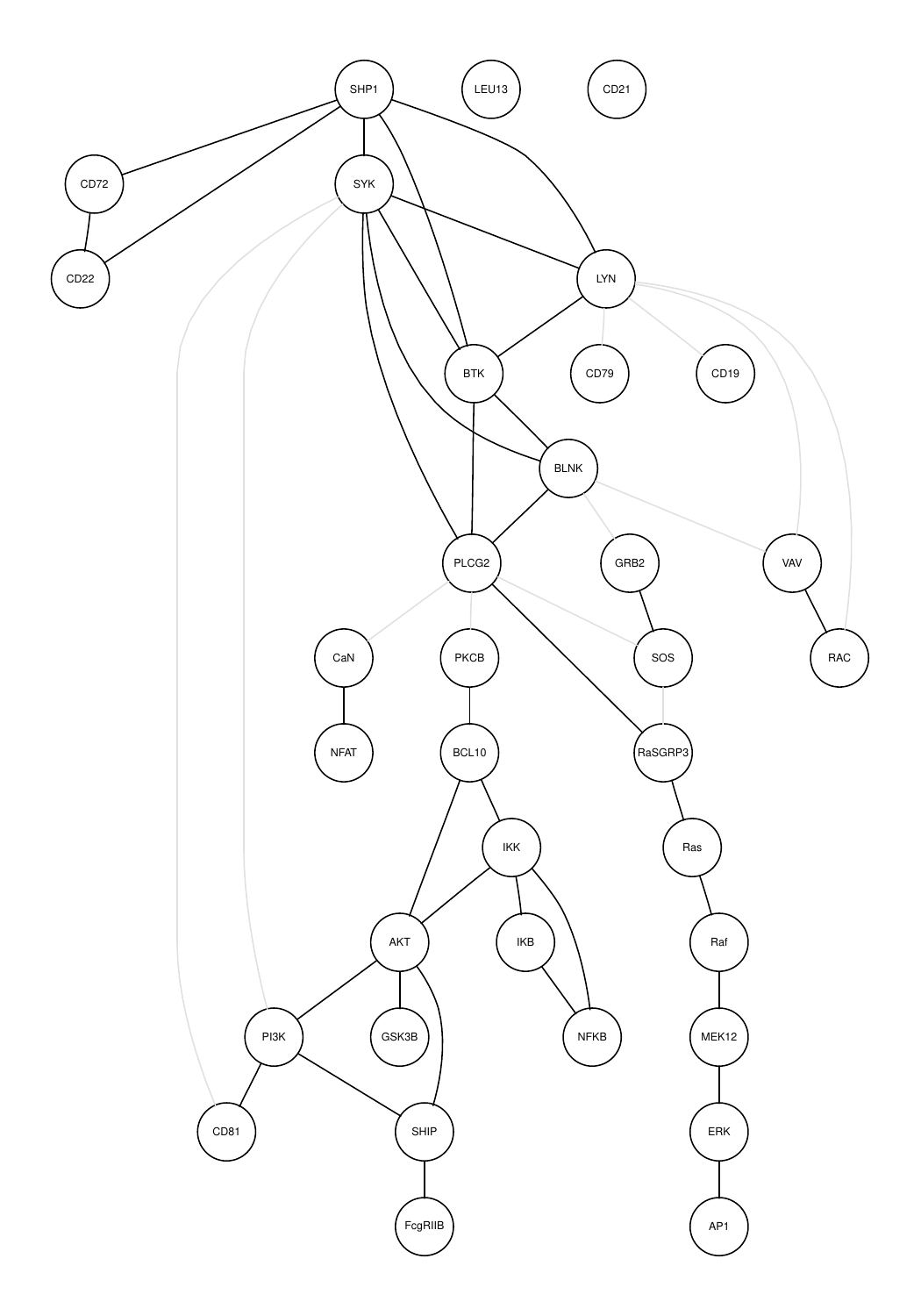}
		\vspace{-0.5cm}
		\caption{BCR signaling pathway involving $q=35$ gene products. The interest is on testing whether a simpler path  without the $d=12$ gray edges can be identified.}
		\label{fig7}
	\end{center}
\end{figure}

\section{Discussion}
\label{sec:disc}

We have provided the theoretical and computational considerations involved in a likelihood-based approach to covariance selection in unsaturated Gaussian graphical models. 
The directional test is based on the saddlepoint approximation to the conditional distribution of sufficient statistics in exponential family models.  The saddlepoint approximation to the conditional density was derived explicitly and proved to be exact within the important class of decomposable models for chordal graphs. Moreover, the computation of the directional $p$-value via one-dimensional numerical integration is made especially fast  by the expedients described in Section \ref{sec:com}. 
Simulations in several scenarios, including situations with a high-dimensional parameter of interest and a large number of nuisance parameters,
illustrate  that the $p$-values from the directional test are  uniformly distributed,
up to the approximation error from the one-dimensional numerical integrations. 
These results provide a confirmation of the theoretical exactness of the saddlepoint-type method with chordal graphs, even if the number of nodes  is greater than the sample size. Our empirical findings suggest also that
the saddlepoint approximation, 
despite not being exact, retains at least the usual accuracy for continuous models when non-chordal graphs are tested.

The likelihood ratio test and its improvements considered here \citep{Skovgaard:2001} are omnibus tests:  the implicit alternative hypothesis is multi-dimensional. In contrast, the directional test uses information in the data to simplify the testing problem to one dimension. The saddlepoint approximation to this distribution incorporates an adjustment for the estimation of the nuisance parameters that has been found to be very effective in simpler problems \citep{pierce92, tang20}.

A natural question about directional tests is whether they entail a loss of power \citep{jensen21}. This is difficult to assess in simulations, as the alternative hypotheses are very high-dimensional. 
We have concentrated in this paper on evaluating the size of the test, which as seen in Tables \ref{tab1}--\ref{tab4} is very well-controlled at conventional $0.05$ and $0.01$ levels, and well into the tails (Figures \ref{fig2}--\ref{fig3}).  We are not aware of any detailed discussions on the power of the likelihood ratio test for these complex Gaussian graphical models with high-dimensional alternatives. 
For high-dimensional normal distributions with $q/n \rightarrow (0,1]$, \citet[Sect. 5.3]{huang21} evaluate the unconditional power of the directional test under a few settings. The performance 
strongly depends on the specific alternative hypothesis under analysis, so it is impossible to draw generally valid conclusions. Still, in those settings the directional test proved to be uniformly more powerful than the likelihood ratio test and its modifications considered here.
It is  also noteworthy that for simpler testing problems in the multivariate normal model, McCormack et al. (2019) showed that the directional test is  equivalent to the uniformly most powerful invariant test based on the $F$ statistic or Hotelling's $T^2$ statistic. 
}


The directional approach detailed here could be extended to graphical models for discrete data, like those dealt with in \cite{roverato17}. 
However, as discreteness prevents the saddlepoint approximation from being exact even upon normalization, one might reasonably not expect the same accuracy of directional $p$-values observed in this work, at least in the most challenging testing problems.

The present methodology only applies to situations where the number of observations is such that the ML estimate exists with probability one under 
the alternative hypothesis. In particular, 
the sample size must be greater than the maximal clique size of the hypothesized graph or its decomposable version \citep{buhl93}. 
The development of reliable likelihood-based testing procedures, omnibus or directional, in circumstances  where the number of nodes is much larger than the number of observations is still an open problem to be addressed in future research.

\section*{Supplementary Materials}
Supplementary materials available at \url{https://github.com/cdicaterina/DirTestGGM.git} provide the data and the R code to reproduce all numerical results in the paper. 
\par
\section*{Acknowledgements}
The authors are grateful to Alberto Roverato for useful discussions and suggestions on the R code to compute the Isserlis matrix. They also thank Davide Risso for his help with the genetic application and Caizhu Huang for suggesting computational improvements.
\par


\appendix
\section*{Appendices}
\renewcommand{\thesubsection}{\Alph{subsection}}
\subsection{Proof of Theorem \ref{th}}  \label{appA}
We want to show that the
saddlepoint approximation 
equals the exact conditional distribution of the sufficient statistic under $H_0$, up to some constant.
The sufficient statistic in our setting is $s=u_k$, i.e. the partition corresponding to the non-zero elements in $\Omega_k$ of $u= n/(n-1)\vech \, \hat{\Omega}^{-1}$ where $\hat\Omega^{-1}=y^\top y/n - y^\top1_n1_n^\top y/n^2$ is the sample covariance matrix.

Substituting in the log-likelihood (\ref{lllk}) the ML and constrained ML estimates of the canonical parameter $\omega_k$ obtained in Section \ref{sec:GM}, we get
\begin{align*}
\exp[\ell(\hat\omega_{0};s) - \ell(\hat\omega_k;s)] &= \left(\dfrac{|\hat\Omega_0|}{|\hat\Omega_k|}\right)^{\frac{n-1}{2}} \exp \bigg[\frac{n-1}{2}\{\hat\omega_k - \hat\omega_{k0}\}^{\top} J_{kk}\hat\sigma_{k} \bigg]\\ 
&= \left(\dfrac{|\hat\Omega_0|}{|\hat\Omega_k|}\right)^{\frac{n-1}{2}}\,,
\end{align*}
since the exponential equals 1 (see Appendix \ref{appB}).
Given equation (\ref{jkk}) in Section \ref{nonsat} for $j_{\omega_k\omega_k} (\omega_k)$, we can then write the expression for the saddlepoint approximation (\ref{conddens}) in our setting as
\begin{equation}\label{app:sadd}
h(s;\psi_0)
\propto
\left(\dfrac{|\hat\Omega_0|}{|\hat\Omega_k|}\right)^{\frac{n-1}{2}} 
|\mathrm{Iss}(\hat\Omega_k^{-1})_{kk}| ^{-1/2}\,.
\end{equation}
Consider now the density of $s=u_k$. This is the marginal density of $p$ entries in $\hat\Omega^{-1}$, the sample covariance matrix with joint Wishart distribution $W_q(n-1, \Omega^{-1}/n)$. Solving the likelihood equation in Section \ref{nonsat} implies that $\hat\sigma_k=u_k=s$, hence these entries are the same as those in the corresponding entries of the matrix $\hat\Omega^{-1}_k$.
We can obtain such a density  for chordal graphs with vertex set decomposable into cliques $C_1, \dots, C_{K}$ and separators $S_2, \dots,S_{K}$ with cardinality $n_{C_i}$ and $n_{S_i}$, respectively. 
Combining the results on the factorization of the joint density of  $\hat\Omega^{-1}$ \citep[(5.45)]{Lauritzen:1996} and on the marginal Wishart distributions for the sub-matrices $\hat\Omega^{-1}_{kC_i}=(\hat\Omega^{-1}_k)_{C_i}$ and $\hat\Omega^{-1}_{kS_i}=(\hat\Omega^{-1}_k)_{S_i}$ \citep[Sect. 7.3.1]{dawid93}, under the null hypothesis $H_0: \omega_k = (\psi, \lambda) = (0, \lambda)$
the true concentration matrix is $\Omega_0$ and so we have:
\begin{align*}
f(s; \Omega_0^{-1}) = & 2^{-\frac{n-1}{2} (\sum_{i=1}^K n_{C_i} - \sum_{i=2}^K n_{S_i})} 
\dfrac{   \prod_{i=1}^K    \Gamma_{n_{C_i}} \left(\frac{n-1}{2}\right)
	|\Omega^{-1}_{0C_i}|^{- \frac{n-1}{2}}
	|\hat\Omega^{-1}_{kC_i}|^{	(n-2-n_{C_i})/2}
}{\prod_{i=2}^K  \Gamma_{n_{S_i}} \left(\frac{n-1}{2}\right)
	|\Omega^{-1}_{0S_i}|^{- \frac{n-1}{2}}
	|\hat\Omega^{-1}_{kS_i}|^{
		(n-2-n_{S_i})/2}
} \\
& 
\cdot \exp\left\{   
-\frac{n}{2} \left[
\sum_{i=1}^K \tr \left(
\hat\Omega^{-1}_{kC_i} \Omega_{0C_i}
\right) 
- 
\sum_{i=2}^K \tr \left(
\hat\Omega^{-1}_{kS_i} \Omega_{0S_i}
\right) 
\right]
\right\}\,.
\end{align*}
Rearranging the factors in the previous formula and neglecting the constants, 
we can write
\begin{align*}
f(s; \Omega_0^{-1}) \propto & 
\left(\dfrac{   \prod_{i=1}^K     |\Omega^{-1}_{0C_i}| 
}{
	\prod_{i=2}^K	|\Omega^{-1}_{0S_i}|
}\right)^{- \frac{n-1}{2}}
\left(
\dfrac{   \prod_{i=1}^K     |\hat\Omega^{-1}_{kC_i}| 
}{
	\prod_{i=2}^K	|\hat\Omega^{-1}_{kS_i}|}
\right)^{\frac{n-1}{2}}
\dfrac{   \prod_{i=1}^K     |\hat\Omega^{-1}_{kC_i}|^{-(n_{C_i}+1)/2}
}{
	\prod_{i=2}^K	|\hat\Omega^{-1}_{kS_i}|^{-(n_{S_i}+1)/2}}	
\\
& 
\cdot \exp\left\{   
-\frac{n}{2} \left[
\sum_{i=1}^K \tr \left(
\hat\Omega^{-1}_{kC_i} \Omega_{0C_i}
\right) 
- 
\sum_{i=2}^K \tr \left(
\hat\Omega^{-1}_{kS_i} \Omega_{0S_i}
\right) 
\right]
\right\}\,.
\end{align*}
We now use the decomposition of the graph \citep[p. 145]{Lauritzen:1996} to re-express the first two factors as a ratio of determinants, the result by \citet{rov98}
mentioned in Section \ref{sec:issdet} to re-express the third factor as the determinant of the Isserlis matrix, and finally the property of the trace operator to re-express the fourth factor. Hence we have
\begin{align*}
f(s; \Omega_0^{-1}) \propto & 
\left(
\dfrac{	|\Omega_0|}{|\hat\Omega_k|}
\right)^{\frac{n-1}{2}}  |\mathrm{Iss}(\hat\Omega_k^{-1})_{kk}| ^{-1/2}\\
&
\cdot \exp\left\{   
-\frac{n}{2} \left[
\sum_{i=1}^K \tr \left(
\Omega_{0C_i}\hat\Omega^{-1}_{kC_i} 
\right) 
- 
\sum_{i=2}^K \tr \left(
\Omega_{0S_i}\hat\Omega^{-1}_{kS_i}
\right) 
\right]
\right\} \\
\propto & 
\left(
\dfrac{	|\Omega_0|}{|\hat\Omega_k|}
\right)^{\frac{n-1}{2}}  |\mathrm{Iss}(\hat\Omega_k^{-1})_{kk}| ^{-1/2}
\exp\left\{   
-\frac{n}{2} \left[
\tr \left(
\Omega_{0} \hat\Omega^{-1}_{k}
\right) \right]\right\}\,,
\end{align*}
where in the last step we have applied again the decomposition property based on the factorization of the density in  chordal graphs \citep[ (5.45)]{Lauritzen:1996} to find the final expression in the exponential of the last factor.
The null conditional density of 
the sufficient statistic in $\mathcal{L}_0$ is given by setting $\omega_k= \hat\omega_{k0}=(0, \hat\lambda_0)$, or equivalently by fixing the concentration matrix under the null hypothesis $\Omega_0$ at its constrained ML estimate $\hat\Omega_0$, i.e.
\begin{align}\label{app:cond}
f(s; \hat\Omega_0^{-1})    \nonumber
\propto  &
\left(
\dfrac{	|\hat\Omega_0|}{|\hat\Omega_k|}
\right)^{\frac{n-1}{2}}  |\mathrm{Iss}(\hat\Omega_k^{-1})_{kk}| ^{-1/2}
\exp\left\{   
-\frac{n}{2} \left[
\tr \left(
\hat\Omega_{0} \hat\Omega^{-1}_{k}
\right) \right]\right\}  \\
\propto  &
\left(
\dfrac{	|\hat\Omega_0|}{|\hat\Omega_k|}
\right)^{\frac{n-1}{2}}  |\mathrm{Iss}(\hat\Omega_k^{-1})_{kk}| ^{-1/2}\,.
\end{align}
In the last step we have used $\tr(\hat\Omega_{0} \hat\Omega^{-1}_{k}) =\tr(\hat\Omega_{0} \hat\Omega^{-1}_0)=\tr(I_q)=q$ (see Appendix \ref{appB}).

Equation (\ref{app:cond}) equals equation (\ref{app:sadd}), up to some constant. 
The normalizing constant of $f(s;\hat\Omega_{0}^{-1})$ simplifies in the ratio of integrals in (\ref{pval}) for computing the directional $p$-value.
The one-dimensional integration is allowed by further restricting on the line $\mathcal{L}^*$ in $\mathcal{L}^0$, identified by $\hat\Omega_k^{-1}(t)= t\hat\Omega_k^{-1} + (1-t)\hat\Omega_0^{-1}$.
As the observed value $\hat\Omega_0$ of the concentration matrix under $H_0$ does not depend on $t$, we can integrate in the numerator and denominator of (\ref{pval}) the function
\begin{equation*}
h(s(t);\psi_0)
\propto
|\hat\Omega^{-1}_k(t)|^{\frac{n-1}{2}} 
|\mathrm{Iss}\{\hat\Omega_k^{-1}(t)\}_{kk}| ^{-1/2}\,,
\end{equation*}
which was given in (\ref{hfun}). 


\subsection{Proof of $\tr[\{ \hat\omega(t) - \hat\omega_{0}\}^{\top} J\hat\sigma(t)  ]=0$}\label{appB}\noindent
We show that the scalar function
\begin{equation*}
f(t) = \{\hat\omega_k(t) - \hat\omega_{k0}\}^{\top} J_{kk}\hat\sigma_{k}(t)
\end{equation*}
equals zero. Since $f(t)=\tr\{f(t)\}$ and the two models under comparison are nested, it is equivalent to prove that $\tr[\{ \hat\omega(t) - \hat\omega_{0}\}^{\top} J\hat\sigma(t)  ]$ is constant in $t$, where
\[ 
\hat\omega(t) = \begin{pmatrix}
\hat\omega_k(t)\\
0
\end{pmatrix}\,,\quad
\hat\omega_0= \begin{pmatrix}
\hat\omega_{k0}\\
0
\end{pmatrix}\,,\quad
\hat\sigma(t)  = \begin{pmatrix}
\hat\sigma_k(t)\\
\hat\sigma_h(t)
\end{pmatrix}\,,
\]
are all vectors of dimension $q^*$. Letting $\hat\Omega^{-1}_k(t)=\Sigma\{\hat\sigma(t)\}$, we have:
\begin{align*}
\tr[\{ \hat\omega(t) - \hat\omega_{0}\}^{\top} J\hat\sigma(t)  ] & = 
\tr[\vech\, \{ \hat\Omega_k(t) - \hat\Omega_{0}\}^{\top} G^{\top}G\, \vech\,\hat\Omega^{-1}_k(t)  ]\\
& = \tr[ \{ \hat\Omega_k(t) - \hat\Omega_{0}\}^{\top} \hat\Omega^{-1}_k(t)  ]\\
& = \tr[ I_q - \hat\Omega_{0} \{ t \hat\Omega^{-1}_k + (1-t) \hat\Omega^{-1}_{0}\}  ]\\
& = \tr(I_q) - t \tr(\hat\Omega_{0} \hat\Omega^{-1}_k)-(1-t) \tr (I_q)\\
& = q - tq - (1-t) q = 0\,.
\end{align*}
This uses basic matrix algebra \citep[see, for instance,][]{abadir05}
and  the equality $\tr(\hat\Omega_{0} \hat\Omega^{-1}_k) =\tr(\hat\Omega_{0} \hat\Omega^{-1}_0)=\tr(I_q)=q$. The latter is due to the fact that the trace of the product of two symmetric matrices is the sum of the element-wise products and, by the ML equation, $\hat\Omega^{-1}_{k}$ differs from $\hat\Omega^{-1}_{0}$ only when the corresponding entries of $\hat\Omega_{0}$ are zero  \citep[cf. also][p. 278]{erik96}.

In order to derive the same result for the scalar
$f(1)=\{\hat\omega_k - \hat\omega_{k0}\}^{\top} J_{kk}\hat\sigma_{k}$, the above calculations can be carried out  imposing $t=1$.

\bibliographystyle{chicago}
\bibliography{dir_ref}

\end{document}